# Peer Pressure Shapes Consensus, Leadership, and Innovations in Social Groups

Ernesto Estrada[*] & Eusebio Vargas-Estrada

Department of Mathematics & Statistics, University of Strathclyde, Glasgow G1 1XH, UK

What is the effect of the combined direct and indirect social influences—peer pressure (PP)—on a social group's collective decisions? We present a model that captures PP as a function of the socio-cultural distance between individuals in a social group. Using this model and empirical data from 15 real-world social networks we found that the PP level determines how fast a social group reaches consensus. More importantly, the levels of PP determine the leaders who can achieve full control of their social groups. PP can overcome barriers imposed upon a consensus by the existence of tightly connected communities with local leaders or the existence of leaders with poor cohesiveness of opinions. A moderate level of PP is also necessary to explain the rate at which innovations diffuse through a variety of social groups.

[*] Correspondence should be addressed to: (email) ernesto.estrada@strath.ac.uk.

The social group's pressure on an individual—peer pressure (PP)—has attracted the attention of scholars in a variety of disciplines, spanning sociology, economics, finance, psychology, and management sciences[1–4]. In analyzing PP we should consider not only those individuals directly linked to a particular person, but also those who exert indirect social influence over other persons as well[5–8]. Although PP is an elusive concept, it can be considered a decreasing function of a given individual's socio-cultural distance from the group. Thus, an individual's opinion may be influenced more strongly by the pressure exerted by those socio-culturally closer to her. Consensus is well documented across the social sciences, with examples ranging from behavioral flocking in popular cultural styles, emotional contagion, collective decision making, pedestrians' walking behavior, and others[9-12].

We can model consensus in a social group by encoding the state of each individual at a given time $t$ in a vector $\mathbf{u}(t)$. The group reaches consensus at $t \to \infty$ when $|u_i(t) - u_j(t)| \to 0$ for every pair of individuals, and the collective dynamics of the system is modeled by

$$\frac{d\mathbf{u}(t)}{dt} = -\mathbf{L}\mathbf{u}(t), \qquad \mathbf{u}(0) = \mathbf{u}_0, \qquad (1)$$

where $\mathbf{L}$ is a linear operator (Laplacian matrix) capturing the topology of the social network[9].

Decisions in groups trying to reach consensus are frequently influenced by a small proportion of the group who guides or dictates the behavior of the entire network. In this situation a group of leaders indicates and/or initiates the route to the consensus, and the rest of the group readily follows their attitudes. The study of leadership in social groups has always intrigued researchers in the social and behavioral sciences[13–17]. Specifically, the way in which leaders emerge in social groups is not well understood[18]. Leaders may emerge either randomly in response to particular historical circumstances or from the individual having the most prominent position (centrality) in the social network at any time.

**Results**
**Emergence of leaders and PP.** To capture the influence of PP over the emergence of leaders in social groups, we consider that the pressure that an individual $p$ receives from $q$ deteriorates proportionally with the social distance between $p$ and $q$. The social distance is captured by the number of links in the shortest path connecting $p$ and $q$. Mathematically, we model the mobilizing power between two individuals at distance $d$ as $\Delta_d \sim f(d)^{-1}$, where $f(d)$ represents a function of the social distance (see Methods equations (11) and (12)). The collective dynamics of the network under peers' mobilizing effects is described by the following generalization of the consensus model

$$\frac{d\mathbf{u}(t)}{dt} = -\left(\sum_d \Delta_d \mathbf{L}_d\right)\mathbf{u}(t), \quad \mathbf{u}(0) = \mathbf{u}_0, \qquad (2)$$

where $\mathbf{L}_d$ captures the interactions between individuals separated by $d$ links in their social network, $\Delta_d \sim 1/d^\alpha$ where the parameter $\alpha$ accounts for the strength of the PP pulling an individual into the consensus.

We now compare the hypotheses about the random emergence of good leaders—those who significantly reduce the time for reaching consensus in a network—to those in which leaders emerge from the most central individuals. Let us examine the emergence of leadership from five centrality criteria: degree, eigenvector, closeness, betweenness, and subgraph (see Supplementary Information equations (S1)-(S5)). In general, we observe that the leaders emerging from the most central individuals are better in leading the consensus than those emerging randomly. However, when there is certain level of PP over the actors, the situation changes dramatically (Fig. 1a, b). First, the time to reach consensus significantly decreases to less than 20% of the time needed when no PP exists. Second, a leader emerging randomly in the network could be as good as one emerging from the most central actors when PP exists in the system. Due to the recent results about the role of low-degree nodes in controlling complex networks[19] we have also tested the role of PP over these potential drivers. Our results show again that good leaders emerge regardless of their centrality in the network when PP exists in the system (Supplementary Information). In other words, under the appropriate PP any individual in a social group could emerge as a good leader independently of her position in the network. This result adds a new dimension to the problem of network controllability[19–22] by demonstrating that PP is a major driving force in determining how potential controllers can emerge in the network independently of their centrality (Supplementary Fig. S1) and — in contrast with previous results[19, 23, 24] — of the degree distribution of the network (Supplementary Fig. S2).

**a**                                              **b**

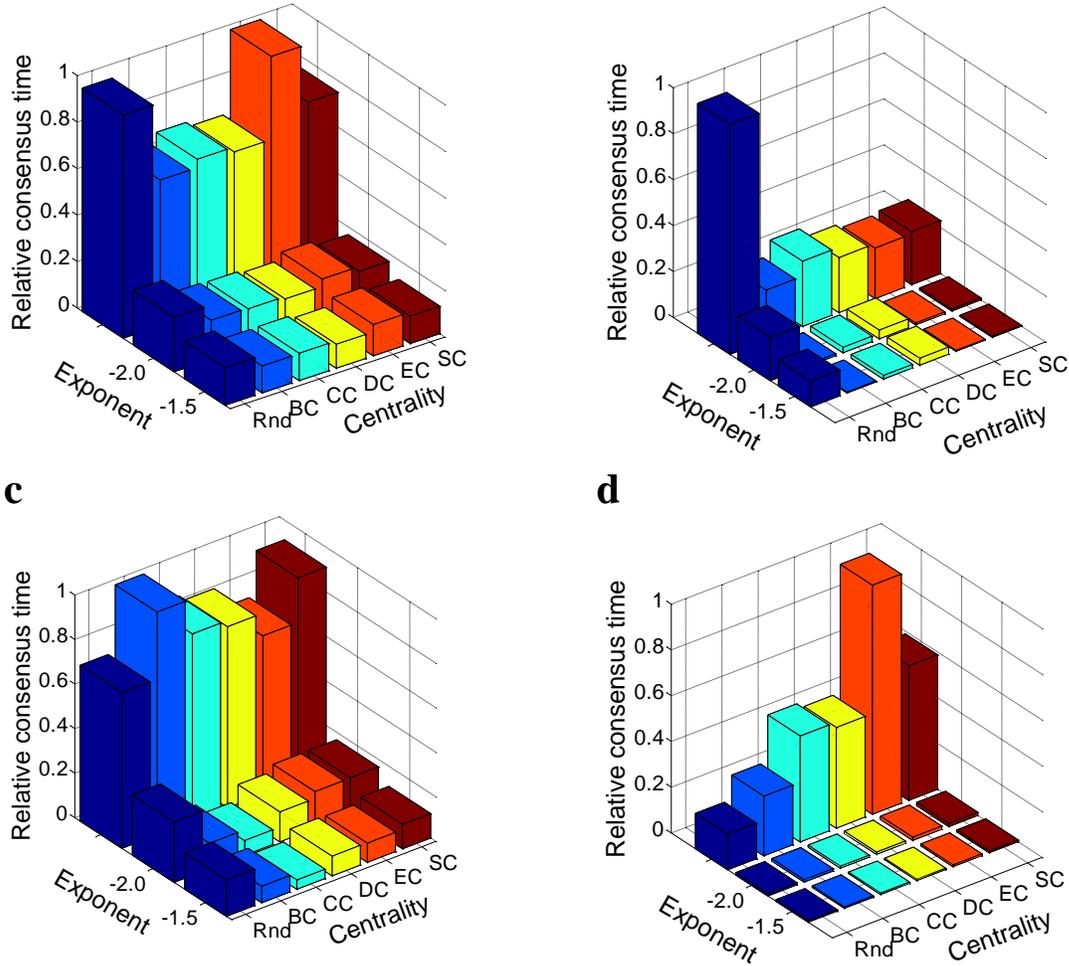

**Figure 1 | Random and centrality-based emergence of leaders**. The emergence of leaders is analyzed according to randomness (Rnd), betweenness (BC), closeness (CC), degree (DC), eigenvector (EC), and subgraph (SC) centrality. The peer pressure is modeled by $\Delta_d \sim 1/d^\alpha$, with $\alpha$ equal to $-1.5$ and $-2.0$. The third line corresponds to no peer pressure. (**a**) Communication network among workers in a sawmill. (**b**) Elite corporate directors. (**c**) Friendship network of injected drug users in Colorado Springs. (**d**) Random network having communities.

In roughly half of the 15 social networks studied (Supplementary Information) we observe the following anomalous pattern. Leaders randomly emerging in the network are better in leading the consensus than some emerging from the most central individuals (see Fig. 1c). This situation appears when the network has the leaders distributed through diverse communities in the network. A *community* is a group of individuals who are more tightly connected among themselves than with the other actors in the network[25]. Actors in one of these communities reach consensus among themselves easily, but it is difficult to reach consensus between different communities. Most central actors in such networks are frequently located in a single community. When they emerge as leaders, they drive consensus only in their community but not in the global network. In contrast, when leaders emerge randomly, they more likely emerge simultaneously in different communities, a situation that favors global agreement in the network. Constructing a random network with communities as illustrated in Fig. 1 (bottom right) corroborates this hypothesis (Supplementary Tables S6 and S7). These results suggest the necessity of considering community leaders in social networks as effective mobilizers of actors throughout the network. We have observed that the leaders emerging on the basis of their community positions exhibit greater success in reaching consensus than those randomly emerging in the network. However, when appropriate PP exists, leaders who effectively reach consensus emerge regardless of their position in their communities.

The leaders in a social group do not always exhibit a high level of cohesiveness. We posit that the leaders' capacity to lead the consensus in a network depends on their divergence of opinions. A cohesive group of leaders can more effectively lead the social group than leaders with larger divergences among their opinions. To model leader cohesiveness we introduce the divergence parameter $\nabla_L$, which is the circumradius of the regular polygon comprising all the leaders. $\nabla_L = 0$ indicates a very cohesive group of leaders. We now examine the influence of the leaders' cohesiveness on consensus. Figure 2 illustrates the results for the friendship network of workers in the sawmill with either no PP (left plots) or with PP modeled by $\Delta_d \sim 1/d^2$ (right plots). The values of leader divergence range from 0.0 to 0.2. The lack of leader cohesiveness significantly increases the time to consensus when there is no PP. In fact, the time increases more than 33% when the divergence changes from 0.0 to 0.2 (it grows to 80.2% for $\nabla_L = 0.5$, see Supplementary Fig. S3 and S4 and Supplementary Tables S1, S3-S5). In addition, the cohesiveness of the group—measured by the standard deviation at consensus $\nabla_G$—is very poor for large values of $\nabla_L$ ($\nabla_G = 154.6$, 183.6, and 226.9 for $\nabla_L = 0.0$, 0.1, and 0.2, respectively), which indicates highly heterogeneous group opinions. However, when PP exists, the situation dramatically changes. First, the time to consensus does not increase as drastically with the decrease of leader cohesiveness. Second, group cohesiveness at the consensus is very high even for the lowest leader cohusiveness ($\nabla_G = 27.0$, 35.4, and 33.0, for $\nabla_L = 0.0$, 0.1, and 0.2, respectively). In short, when PP is absent, leader cohesiveness plays a fundamental role in the time needed to reach consensus and in group cohesiveness at the consensus. When PP is present, the time needed to reach consensus and group cohesiveness are largely independent of the degree of divergence in the leaders' opinions, and the consensus is driven primarily by the influence of the nearest neighbors and PP.

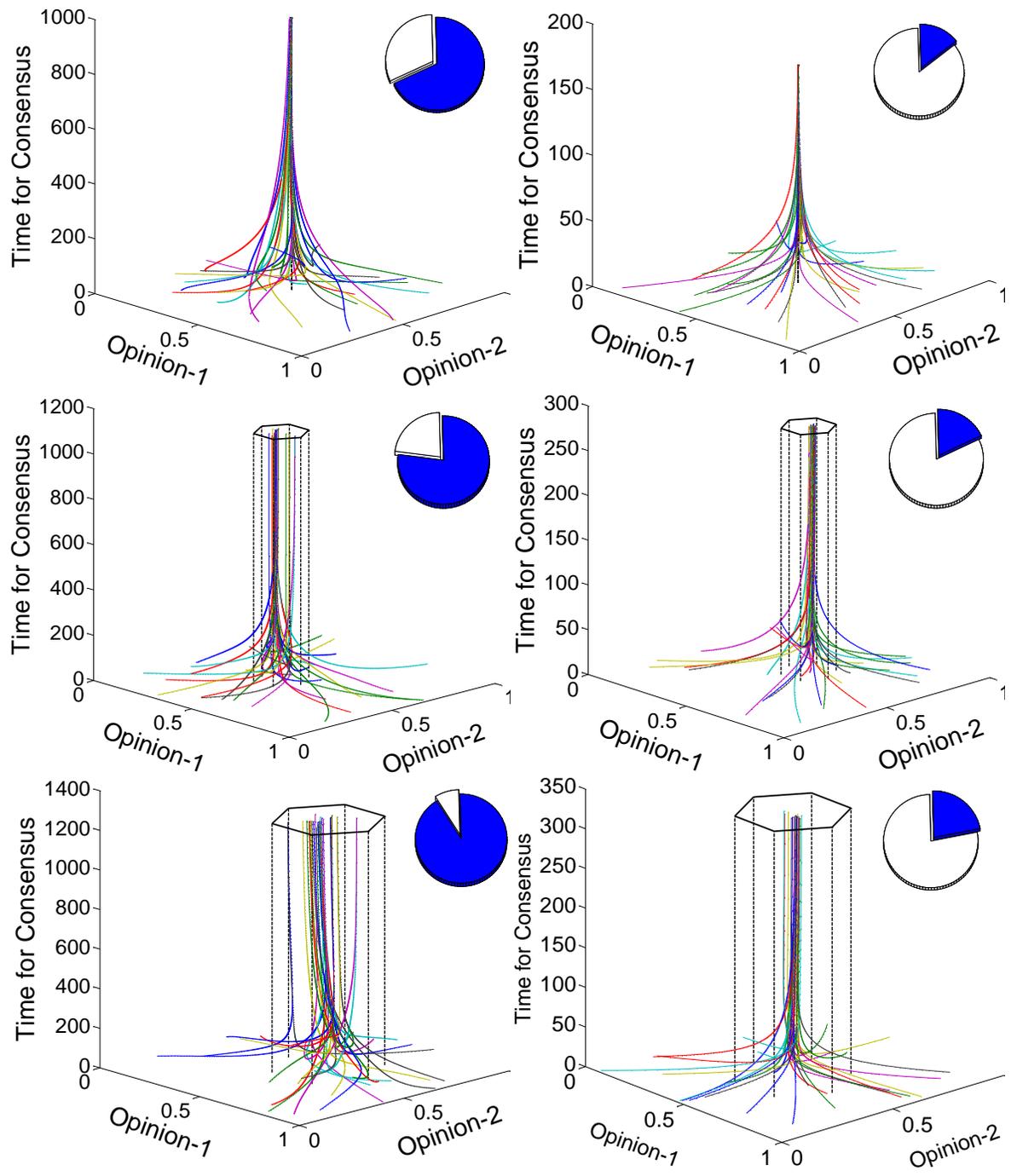

**Figure 2 | Leaders' cohesiveness and consensus**. Analysis of the influence of leaders cohesiveness on the time to reach consensus in the communication network among workers in the sawmill without (left plots) and with (right plots) PP. The leaders' divergences used are: 0.0 (top), 0.1 (middle), and 0.2 (bottom). The time to reach consensus (in blue) relative to a total time of 1,500 units (Insets).

**Diffusion of innovations and PP.** Another area that has received great research attention is the diffusion of innovations[26–29]. The diffusion of innovations refers to the process through which new ideas and practices spread within and between social groups. Here we consider the hypothesis that PP plays a fundamental role in innovation adoption or rejection. To test our hypothesis, we study two datasets in which diffusion of innovations was followed for different periods of time (Supplementary Information). The first study analyzed the diffusion of a modern mathematic method among the primary and secondary schools in Allegheny County (Pennsylvania, USA). Results revealed that innovation diffused through the friendship network of the superintendents of the schools involved. The study was followed for a period of six years, 1958–1963. The second dataset represents the second phase of a longitudinal study about how Brazilian farmers adopted the use of hybrid seed corns, examining personal factors influencing farmers' innovative behavior in agriculture. We consider here the social network of friendship ties and the cumulative number of adopters of the new technology in three different communities of the Brazilian farmers study (Supplementary Fig. S5). The study was conducted over the course of 20 years and we consider only the individuals in the largest connected components of the networks.

Figure 3 depicts the number of actors that adopted the respective innovations at different times. These values correspond to the number of adopters observed empirically in field studies. To simulate the process of innovation adoption, we study the consensus dynamics with equation (2), assuming $\Delta_d \sim 1/d^\alpha$: no PP, moderate PP ($-6.0 \leq \alpha \leq -5.0$), high PP ($-4.0 \leq \alpha \leq -3.0$) (see Supplementary Information). The simulations follow perfect sigmoid curves, as Fig. 3 illustrates. Observe that when there is no PP effect, the diffusion curves predict slower rates of adoption than those empirically observed. For example, the empirical evidence demonstrates that 50% of schools adopted the new math method in roughly three years, whereas the simulation without PP predicts a period of four years of a total of six years. In the case of the Brazilian farmers, the empirical time for 50% of the farmers to adopt the innovation is roughly 12 years, whereas the simulation without PP predicts 16 years of a total of 20 years. When the model uses strong PP, the diffusion curves display very rapid adoption rates, which are far from the reality of the empirical evidence in both cases. However, using a moderate PP predicts very well the outputs of the empirical results in both studies. These PP values are found by a reverse engineering method, but the important message is that a certain PP level is necessary to describe the empirical evidence on the diffusion of innovations in social groups (see also Supplementary Information).

These results demonstrate that interpersonal communication alone cannot sufficiently explain the process of innovation adoption in a social group. The pressure exerted by the social group plays a fundamental role in shaping this important social phenomenon. Our model describes effectively PP's role in these and other important phenomena, consistent with our intuition and with the existing empirical evidence.

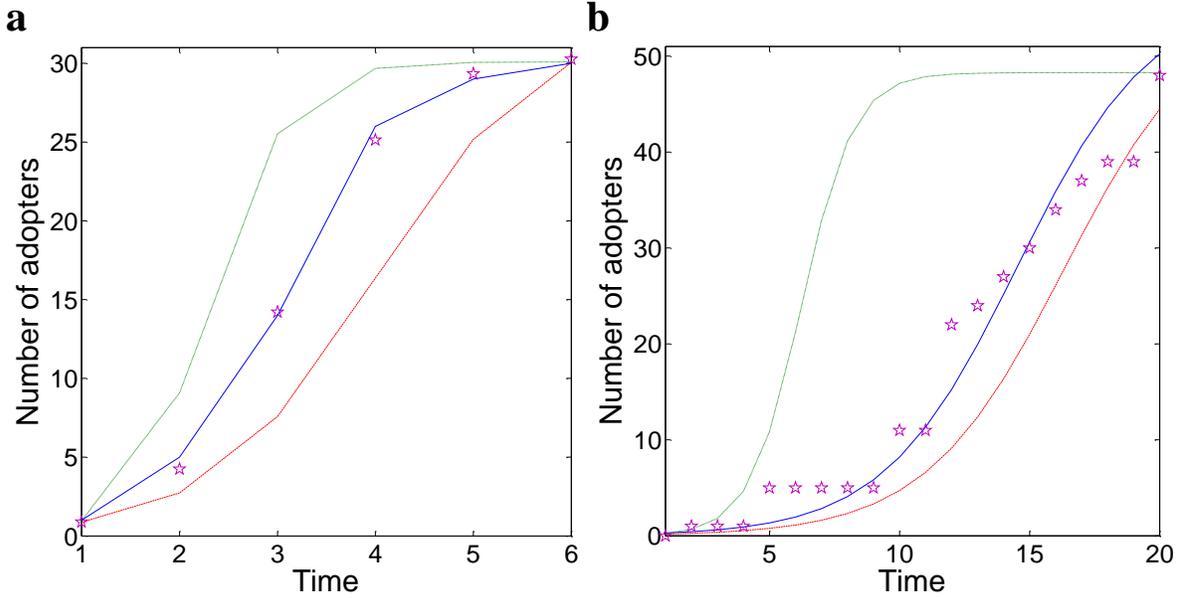

**Figure 3 | Diffusion of innovations under PP.** (**a**) Adopters of a new mathematical method among US colleges in a period of 6 years. (**b**) Adopters of the use of hybrid seed corns among Brazilian farmers for a period of 20 years. Experimental values are given as stars and the simulation with no (broken red line), moderate (continuous blue line) and strong (dotted green line) PP are illustrated.

**Methods**

**Consensus dynamics model.** We consider a social group of $n$ actors who will accomplish a certain goal or reach an agreement. Every actor in the group is represented by an element of the node set $V = \{1,...,n\}$ of a network $G = (V, E)$, in which links (edges) $E \subseteq \{V \times V\}$ represent the relationships (friendship, any form of communication) among the actors. The set of neighbors of the actor $i$ is denoted by $N_i = \{j \in V : (i, j) \in E\}$. Let $\mathbf{A} = [a_{ij}] \in R^{n \times n}$ and $\mathbf{L}(G) = [l_{ij}] \in R^{n \times n}$ be the adjacency matrix and Laplacian matrix, respectively, associated with graph $G$. The Laplacian matrix is defined as $\mathbf{L} = \mathbf{K} - \mathbf{A}$, where $\mathbf{K}$ is the diagonal matrix of node degrees of $G$ and $\mathbf{A}$ is the adjacency matrix.

The information states of the actors evolve according to the single-integrator dynamics given by

$$\frac{du_i(t)}{dt} = g_i, \ i = 1,...,n, \text{ and } u_i(0) = z_i, \quad (3)$$

where $u_i \in R$ is the information state at time $t$, $g_i \in R$ is the information control input, and $z_i \in R$ is the initial state of actor $i$, which is always considered to be selected at random. A continuous time consensus algorithm is given by

$$g_i = \sum_{j \in N_i} a_{ij}(u_j(t) - u_i(t)), \ i = 1,...,n, \quad (4)$$

where $a_{ij}$ is the $(i, j)$ entry of the adjacency matrix $\mathbf{A}$. The information state of each actor is driven toward those of her neighbors. Equations (3) and (4) describe the collective dynamics of the social group and can be written in matrix form as

$$\frac{d\mathbf{u}(t)}{dt} = -\mathbf{Lu}, \quad (5)$$

where $\mathbf{u} = [u_1,...,u_n]^T$ is the vector of the states of the actors in the system. The consensus among the actors is achieved if, for all $u_i(t)$ and all $i, j = 1,...,n$, $|u_i(t) - u_j(t)| \to 0$ as $t \to \infty$.

When the interaction among agents occurs at a discrete time, the information state is updated using a difference equation, and a discrete time consensus algorithm is then given by

$$u_i(t+1) = u_i(t) + \varepsilon \sum_j a_{ij}(u_j(t) - u_i(t)), \ i = 1,...,n, \quad (6)$$

where $a_{ij}$ is as before and $\varepsilon$ is the time step. The information state of each actor is updated as the weighted average of her current state and those of her neighbors. Equation (6) is written in matrix form as

$$\mathbf{u}(t+1) = \mathbf{P}\mathbf{u}(t). \tag{7}$$

The matrix $\mathbf{P}$ is known as the Perron matrix, which is obtained as $\mathbf{P} = \mathbf{I} - \varepsilon \mathbf{L}$, for $0 < \varepsilon < 1/\kappa_{max}$, where $\kappa_{max}$ is the maximum of the degrees of the nodes of $G$. The entries of the Perron matrix satisfy the property $\sum_j p_{ij} = 1$ with $p_{ij} \geq 0, \forall i,j$, and hence, it is a valid transition matrix[9].

**Consensus with leaders–followers.** We consider that there exist one or multiple leaders who guide the entire group to the consensus through the effect produced by the rest of the group, which follows them[30]. In a leaders–followers structure with a single leader, actors attempt to reach an agreement that is biased to the state of the leader, whereas in the case of multiple (stationary) leaders, all followers converge to the convex hull formed by the leaders' states.

An actor is called a stationary *leader* if her opinion is available for the other actors but is not modified during the process. Then, the set of all actors can be divided into two subgroups: leaders and followers. As a result, the vector of the states of all actors can also be divided into two parts: the states of leaders, $u_l$, and the states of followers, $u_f$.

For a system with multiple stationary leaders, all the nodes can be labeled such that the first $n_f$ represents the followers and the remaining $n_l$ represent the leaders. The total number of actors in the system is $n = n_f + n_l$, such that the Laplacian matrix associated with the social network $G$ is partitioned as

$$\mathbf{L}(G) = \begin{bmatrix} \mathbf{L}_f & \mathbf{l}_{fl} \\ \mathbf{l}_{fl}^T & \mathbf{L}_l \end{bmatrix}, \tag{8}$$

where $\mathbf{L}_f \in R^{n_f \times n_f}$, $\mathbf{L}_l \in R^{n_l \times n_l}$, and $\mathbf{l}_{fl} \in R^{n_f \times n_l}$.

Because the leaders are stationary, their dynamics are given by $u_i(t) = 0, i = n_f + 1,..,n$. Then, the dynamics of the system are expressed by

$$\begin{bmatrix} \dot{\mathbf{u}}_f \\ \dot{\mathbf{u}}_l \end{bmatrix} = -\mathbf{L}_p \mathbf{u} = -\begin{bmatrix} \mathbf{L}_f & \mathbf{l}_{fl} \\ \mathbf{0} & \mathbf{0} \end{bmatrix} \begin{bmatrix} \mathbf{u}_f \\ \mathbf{u}_l \end{bmatrix}. \tag{9}$$

The discrete version of equation (9) is given by

$$\mathbf{u}(t+1) = (\mathbf{I}_n - \varepsilon \mathbf{L}_p)\mathbf{u}(t), \tag{10}$$

where $\mathbf{u}(t) = [u_1(t),...,u_n(t)]^T$, $\mathbf{I}_n$ is the identity matrix of size $n \times n$, and $\mathbf{L}_p$ is the Laplacian matrix of network $G$, with each entry of the $j$th row equal to zero for $j = n_f + 1,...,n$.

**Modeling peer pressure.** The consensus dynamic modeling assumes that the actors only interact with their directly connected neighbors to cooperatively achieve an agreement in the system[31]. However, in many real-world situations, the actors are exposed not only to their closest contacts but also to individuals who are socio-culturally close to them despite not being directly connected. For instance, this situation appears in actors' attitudes toward copying others. The predisposition of an actor to copy a behavior depends not only on her friends' adoption of such behavior but also on other, socio-culturally close people having a positive predisposition to that behavior. For instance, adolescents adopt "binge drinking" not only by copying their mates but also by observing similar behavior among others of a similar age, education, and social class. Then, we argue that this socio-cultural distance can be captured in a model by considering the shortest path distance between two actors in their social group. The shortest path distance is the number of steps in the shortest path connecting the two actors. The influence that an actor receives/produces from/for others in her social network, i.e., peer pressure, decays as a function of this socio-cultural distance, which separates the two actors[32].

Peer pressure can then be modeled by considering the generalized Laplacian matrix[33]. Consequently, the consensus dynamics model of equation (6) can be written as

$$\mathbf{u}(t+1) = \left(\mathbf{I}_n - \varepsilon\left(\sum_d \Delta_d \mathbf{L}_d\right)\right)\mathbf{u}(t),  \qquad (11)$$

where $\sum_d \Delta_d L_d$ involves the $d$-Laplacian matrices and the coefficients $\Delta_d$ indicate the strength of the interactions at distance $d \leq d_{\max}(G)$, with $d_{\max}(G)$ being the maximum distance between two nodes or the diameter of graph $G$. The $d$-Laplacian matrix is defined as[33]

$$L_d(i,j) = \begin{cases} -1 & d_{ij} = d \\ \upsilon_d(i) & i = j \\ 0 & otherwise \end{cases}, \tag{12}$$

where the expression $\upsilon_d(i)$ is the *d*-path degree of node *i* defined as the number of non-redundant shortest paths of length *d* having *i* as an endpoint.

The coefficients $\Delta_d$ should account for the decay in peer pressure for the socio-cultural distance between the actors of $\Delta_d \sim f(d)^{-1}$, where $f(d)$ represents a function of distance *d*. In this study, we consider three different decay behaviors described by the following equations:

1) Power-law decay: $\Delta_d = d^{-\alpha}$,

2) Exponential decay: $\Delta_d = e^{-\beta d}$, and

3) Social interactions: $\Delta_d = d\delta^{d-1}$,

where α, β, and δ are parameters to be adjusted to consider the different strengths of peer pressure.

# Supplementary Information

**This PDF file includes the following information:**

    Supplementary Methods and Discussion
    Supplementary Dataset Summary
    Supplementary Figures S1–S3
    Supplementary Tables S1–S39
    Supplementary References

**Supplementary Methods and Discussion**

**Emergence of leaders.** We consider two possible scenarios that describe the manner in which leaders emerge in a social group:

1) Emerging randomly from the social group, and
2) Emerging from the most central individuals in the social group.

The centrality of an actor in her social group can be considered in several ways. The concept of actor centrality is related to the question "Which are the most *important* or *central* nodes in a network?" We study five centrality measures defined in[25] to consider potential leaders from among the actors in specific respective social groups. The centrality measures considered are as follows:

i) *Degree centrality* (DC): This measure is considered the simplest in a network defined as the number of edges connected to a node. It has been used assuming that nodes with connections to many other nodes might have more influence or access to information than those with few connections.
The degree of a node can be expressed in a matrix as

$$\kappa_i = (\mathbf{A}\mathbf{1})_i, \quad (S1)$$

where **1** is a column vector of ones and **A** is the adjacency matrix of the network.

ii) *Eigenvector centrality* (EC): This measure appears as an extension of the degree of centrality. Eigenvector centrality is based on the question that not all neighbors are equivalent because, in some cases, the importance of a node is related to (and increased by) its neighbors, which may themselves be important. Thus, instead of giving only one point for each neighbor, this measure gives each node a score proportional to the sum of its neighbor's scores.
The eigenvector centrality of node $i$ is given by the $i$th entry of the principal eigenvector of the adjacency matrix

$$\varphi_i(i) = \left(\frac{1}{\lambda_1}\mathbf{A}\varphi_1\right)_i. \quad (S2)$$

iii) *Closeness centrality* (CC): This index measures the inverse of the mean distance from a node to other nodes and characterizes the nodes according to their distance to all other nodes in the network. The closeness is defined as

$$CC(u) = \frac{n-1}{s(u)}, \quad (S3)$$

where $s(u) = \sum_{v \in V(G)} d(u,v)$ is the distance sum of node $u$.

iv) *Betweenness centrality* (BC): This concept measures the extent to which a node lies on paths between other nodes. The nodes with high betweenness centrality may have considerable influence within a network because of their control over information passing through them.
The index can be defined as

$$BC = \sum_i \sum_j \frac{\rho(i,k,j)}{\rho(i,j)}, i \neq j \neq k, \quad (S4)$$

where $\rho(i,j)$ is the number of shortest paths from node $i$ to $j$, and $\rho(i,k,j)$ is the number of these shortest paths that pass through node $k$.

v) *Subgraph centrality* (SC): This measure is based on the notion that the importance of a node can be characterized by considering its participation in all closed walks for which it is the starting point.
Subgraph centrality has been defined as

$$EE(i) = \left( \sum_{l=0}^{\infty} \frac{\mathbf{A}^l}{l!} \right)_{ii} = \left( e^{\mathbf{A}} \right)_{ii}. \quad (S5)$$

Due to recent discoveries[19] about the role of nodes with low degree as controllers or drivers in complex networks we have also studied the role of PP when the leaders emerge from nodes with high, medium and low degree. In the Supplementary Figure S1 we illustrate the results for two social networks which show that the effect of PP over the emergence of leaders is independent of the status of the nodes in their complex networks. In addition we display in Supplementary Figure S2 the cumulative degree distributions for all the social networks studied here. Our results resumed in Supplementary Tables S12-S39 show that the effect of PP is also independent of the degree distribution of the networks.

**Divergence of leaders' opinions.** We consider that leaders' opinions may differ from the group's average opinion. We call this difference the divergence $\nabla_L$, which is represented by the circumradius of the regular polygon (for a two-dimensional case) that covers all opinions of the leaders (see Supplementary Fig. S3). If the concerned problem is multidimensional, that is, if we are considering more than two opinions in the system, then the divergence is the circumradius of a hypersphere that covers the opinion of all leaders.

We applied our model to two simple undirected random graphs and 15 real social networks described in the Dataset summary. We allowed for every actor to have two opinions that were considered independent from each other, enabling a two-dimensional decoupled consensus process. The initial states of the followers were randomly assigned for every process with values in the range (0, 1). We consider that only six leaders emerge, either randomly or from among the most central actors, who are determined according to their centralities values. Their initial positions were assigned to have divergences of 0.1 and 0.2 from the average consensus. The consensus dynamics stop when the difference between two consecutive measures of disagreement is less than or equal to $1e-07$. We simulate consensus processes with and without PP. The normalized values of every network are reported in Supplementary Tables S12–S39. Every value is the average of 50 realizations.

To test the effects of the divergence in the system dynamics, we used the social network Sawmill and simulated a consensus process that allowed for different PP intensities. We increased the value of divergence on leaders' opinions from zero to a maximum of 0.5. We indicate the consensus times for all divergence values on Supplementary Tables S2–S5. For the case of randomly selected leaders, times increased along with divergence from 13.18% to 80.24% (see Supplementary Table S1).

We highlight that PP influenced the trajectories of followers' opinions, precisely directing them toward the consensus space. At the consensus point, the final positions were more cohesive (Supplementary Fig. S4), indicating more homogeneous final opinions in the system.

**Networks with communities.** We discovered (see main text) that when leaders are spread among different tightly connected communities, "anomalous" patterns are observed in their emergence. This "anomaly" lies in the fact that leaders emerging randomly in the network are more efficient in reaching consensus than those emerging from among the most central actors. We tested this hypothesis by constructing random networks in which we control the number of communities as well as the connectivity within and among each community. These random networks with communities comprised simple, undirected random graphs with 500 nodes and 10 communities generated using the benchmarks for community detection in[34] coded in the C programming language.

First, we selected 10 leaders randomly and by the global highest centralities and recorded the average time for consensus of 50 realizations. We allowed for the emergence of 10 leaders by the local (community-based) highest centrality, i.e., one leader in each community corresponding to the one with the highest centrality. When the leaders emerged from the global highest centralities, we observed the previously described anomalous behavior. This situation was modified when the leaders emerged from the community-based centrality. In this case, the leaders emerging from the most central local leaders were significantly better at reaching consensus than those emerging randomly (see Supplementary Tables S6 and S7).

**Diffusion of innovations.** The essence of the diffusion process is the information exchange by which one individual communicates a new idea to one or several others; thus, diffusion is a special type of communication concerned with the spread of messages perceived as new. In its most elementary form, the main elements in the diffusion process are[26]:

1) An innovation (message or information);
2) Communication channels, through which messages are conveyed from one individual to another;
3) Time of diffusion; and

4) The social system through which the process occurs—a set of interrelated units engaged in joint problem solving to accomplish a common goal. The members or units of a social system may be individuals, informal groups, organizations, and/or subsystems. All members cooperate at least to the extent of seeking to solve a common problem. This common objective binds the system.

Most innovations have a sigmoid-shaped (S-shaped) rate of adoption. The slope of the curve varies with every innovation. Certain new ideas diffuse relatively rapidly, and its S-curve is quite steep. Other innovations may have a slower rate of adoption, reflected by a more gradual S-curve. This behavior indicates the rate of adoption, i.e., the number of adopters of the new idea throughout time. The behavior of the adopters builds the form of the innovation process as follows[26]: Initially, few individuals adopt the innovation in each period; these are the innovators. Soon, the diffusion curve begins climbing as an increasing number of individuals adopt. Then, the trajectory of the rate of adoptions begins to level off, as fewer individuals remain who have not yet adopted. Finally, the S-curve reaches its asymptote and the diffusion process is complete.

Most individuals do not evaluate an innovation on the basis of scientific studies regarding its consequences; most people primarily depend on a subjective evaluation from individuals such as themselves who previously adopted the innovation, i.e., a dependency on the communicated experience of near peers.

In diffusion networks, certain individuals play different roles in a social system, and these roles affect diffusion. Certain members of the system function as opinion leaders: individuals who can influence others, and who are often identified and used to assure better diffusion of the information.

To analyze the impact of PP on the diffusion of innovations process, we used three networks from two empirical studies:

1) The network from the study Mathematical Method[35]: This innovation concerns the diffusion of a new mathematics method in the late 1950s. It was instigated by top mathematicians and sponsored by the U.S. National Science Foundation and the U.S. Department of Education. The diffusion process was successful because most schools adopted the new method. The example traces the diffusion of the modern mathematical method among school systems that combine elementary and secondary programs in Allegheny County (Pennsylvania, U.S.). All school superintendents who were in office for at least two years were interviewed.

   Among other things, the superintendents were asked to indicate their friendship ties with other superintendents in the county through the following question: Among the chief school administrators in Allegheny County, who are your three best friends?

   The researcher analyzed the friendship choices among superintendents who adopted the method and who were in office for at least one year before the first adoption, indicating that they could have adopted earlier. Unfortunately, the researcher did not include the friendship choices by superintendents who did not receive any choices themselves.

In our study, the network represents the friendship ties among the 30 superintendents who were part of the connected component, and the times for adoption represent the year in which the adopter chose the new mathematical method: 1-1958, 2-1959, 3-1960, 4-1961, 5-1962, and 6-1963.

2) Three networks from the study Brazilian Farmers[36]: "Diffusion and Adoption of Innovations in Rural Societies, 1952–1973," was a longitudinal study on how Brazilian farmers (BF) adopted hybrid seed corns. The study was part of a broader, three-phase research project concerned with the spread of modern technology in Brazil, Nigeria, and India.

The data files reflect the second phase, which examined personal factors influencing farmers' innovative agricultural behavior. Villages were selected from the total sample of Phase II villages. The groups of people were divided into different communities according to different variables, and the social networks of friends among the people in each community were retrieved.

The data used for our study includes the social network of friendship ties and the cumulative number of adopters of the new technology over 20 years among the individuals in the giant connected component for three different communities of the study, identified as communities 23, 70, and 71[37].

We applied our consensus model to these networks, and the average consensus time of 50 realizations was divided into six intervals for the Mathematical Method and 20 intervals for the BF networks. We counted the number of actors or nodes corresponding to the average consensus at every time step by measuring the difference between a node's position and the average consensus. When the absolute value of this difference was less than or equal to 0.04, we considered the node to be in agreement. This process was conducted with and without PP by considering a power-law decay.

The cumulative average of nodes in agreement at every interval is shown in Supplementary Tables S8–S11 for every network. In addition, the empirical cumulative number of adopters is indicated. We vary the values of parameter α to obtain behaviors that more effectively follow the empirical patterns. We divided these values into two classes: moderate PP ($-6.0 \leq \alpha \leq -5.0$) and high PP ($-4.0 \leq \alpha \leq -3.0$). Supplementary Figure S5 illustrates the curves of the diffusion processes that indicate that all results have sigmoid-like behavior that varies according to the PP in the system.

**Supplementary Dataset Summary**

| Name | n | Description (23) |
|---|---|---|
| ER | 150 | Simple undirected random graph generated from the Erdös–Rényi model implemented in the toolbox CONTEST (Taylor and Higham, 2008). |
| BA | 150 | Simple undirected random graph generated from the preferential attachment model implemented in the toolbox CONTEST (Taylor and Higham, 2008). |
| Prison | 67 | Social network of prison inmates who chose "Which fellows on the tier are you closest friends with?" (MacRae, 1960) |
| Dolphins | 62 | An undirected social network of frequent associations between 62 dolphins (bottlenose) in a community living in New Zealand (Lusseau, 2003). |
| HS | 69 | Network of relations in a high school. The students choose the three members they wanted to have in a committee (Zeleny, 1950). |
| Zachary | 34 | Data collected by Wayne Zachary from the members of a university karate club, representing the presence or absence of ties among club members (Zachary, 1977). |
| Sawmill | 36 | A communication network within a small enterprise. All employees were asked to indicate the frequency with which they discussed work matters with each of their colleagues on a five-point scale ranging from less than once a week to several times a day (Michael and Massey, 1997; de Nooy *et al.*, 2005). |
| High Tech | 33 | A small high-tech computer firm that sells, installs, and maintains computer systems. The network contains the friendship ties among the employees (Krackhardt, 1999; de Nooy *et al.*, 2005). |
| Galesburg | 31 | From the Columbia University Drug Study. The diffusion of a new drug (gammanym) was investigated and the friendship ties among 31 physicians were coded (Coleman *et al.*, 1966; Knoke and Burt, 1983; de Nooy *et al.*, 2005). |
| Corporate | 1586 | American corporate elite, formed by directors of the 625 largest corporations, that reported the compositions of their boards, selected from the *Fortune* 1,000 in 1999 (Davis *et al.*, 2003). |
| Drugs | 616 | Social network of injecting drug users (IDUs) who shared a needle in the last six months (Moody, 2001). |
| Colorado Springs | 324 | The risk network of persons with HIV during its early epidemic phase in Colorado Springs, U.S., selected through analysis of community-wide HIV/AIDS contact tracing records during 1985–1999 (Potterat *et al.*, 2002). |
| Math Method | 30 | This network concerns the diffusion of a new mathematics |

|  |  | method in the 1950s. It traces the diffusion of the modern mathematical method among school systems that combine elementary and secondary programs in Allegheny County (Pennsylvania, U.S.) (de Nooy *et al.*, 2004). |
|---|---|---|
| Social3 | 32 | Network of social contacts among college students participating in a leadership course. The students choose the three members they wished to include in a committee (Zeleny, 1950). |
| BF23, BF70, BF71 | 40, 48, 49 | Networks of friendship ties from the communities identified as 23, 70, and 71 from the Brazilian Farmers longitudinal study on the adoption of a new corn seed (Valente, 2012; Herzog *et al.*, 1968). |

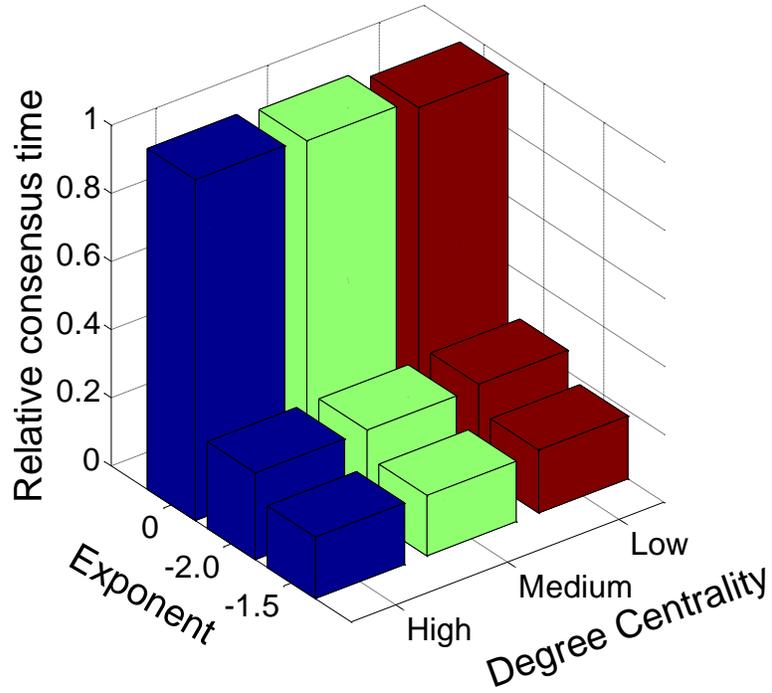

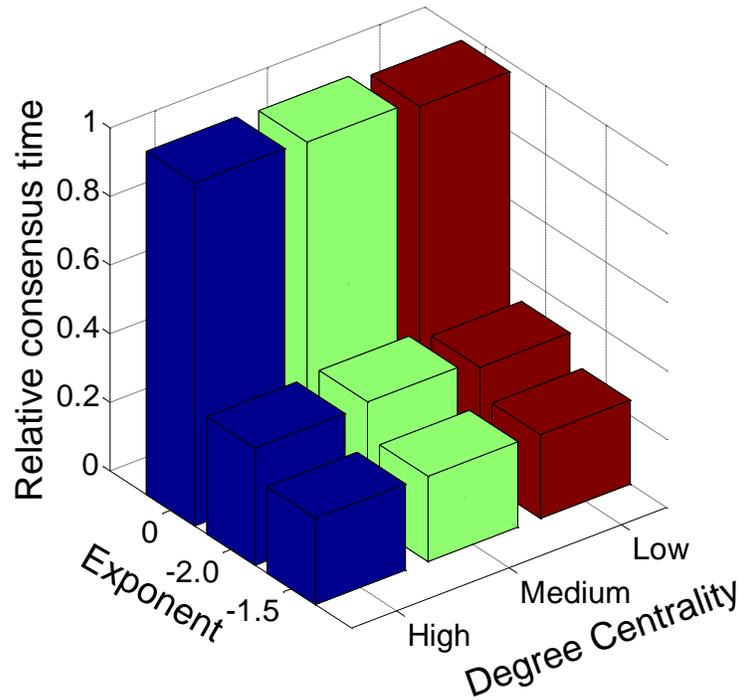

**Supplementary Figure S1 | Influence of PP on emergence of leaders by selecting them among the nodes with high, medium or low degree centrality**. The HighTech (**a**) and the network of social dates (social3) (**b**).

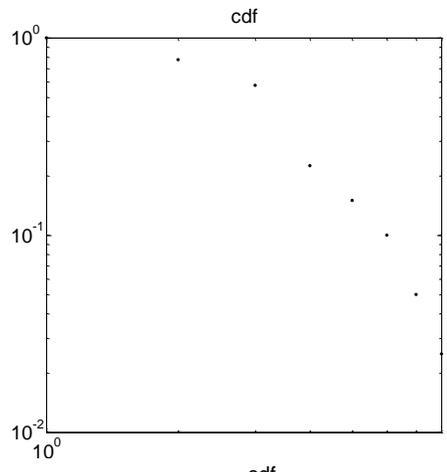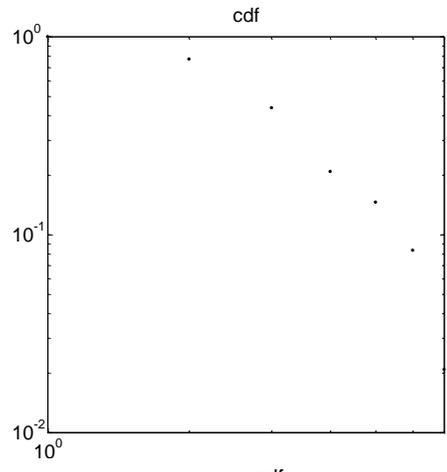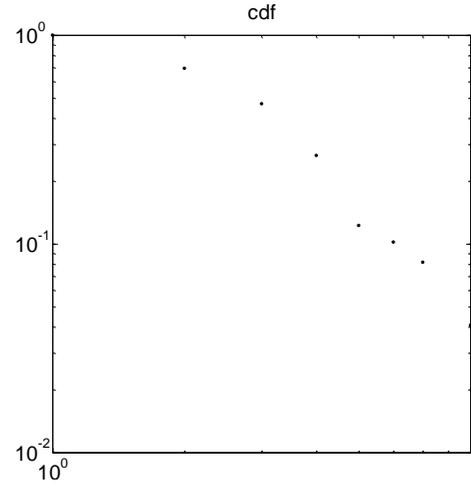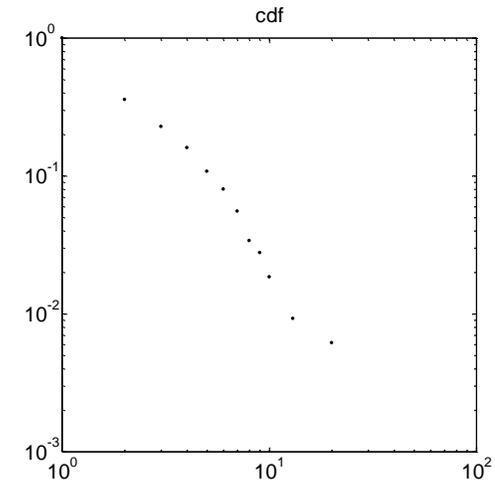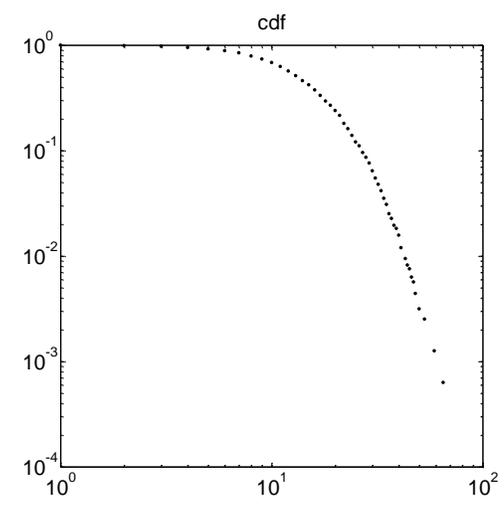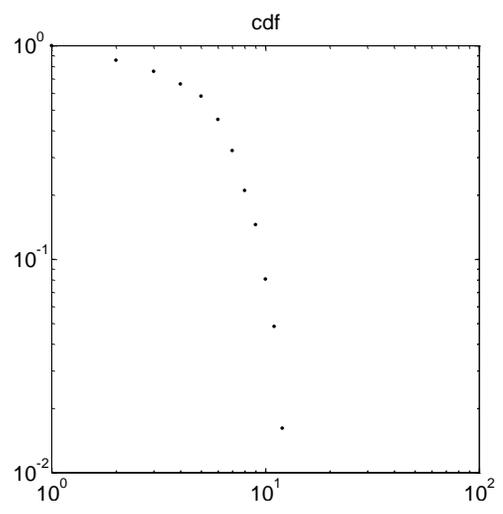

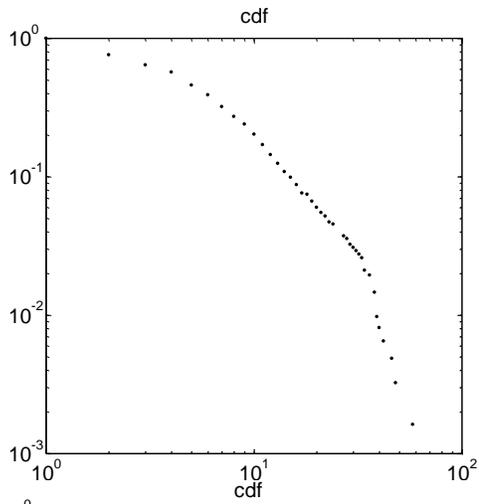
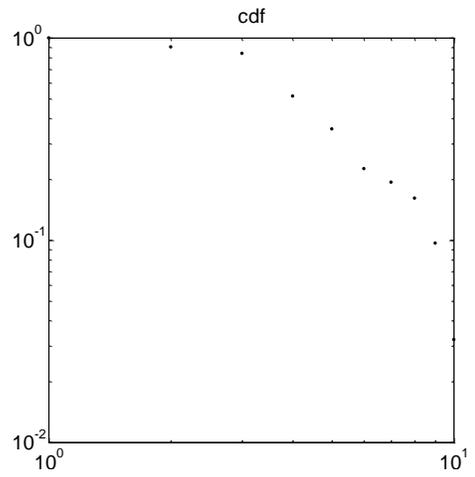
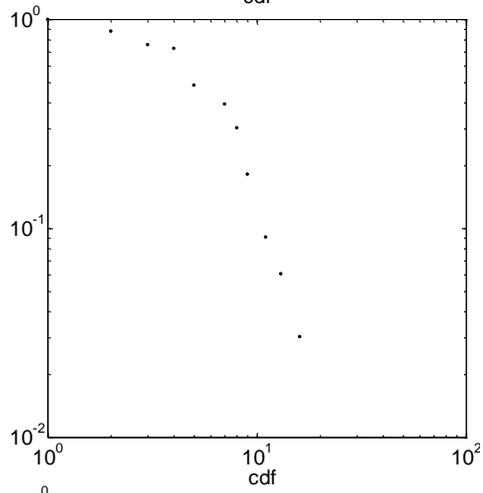
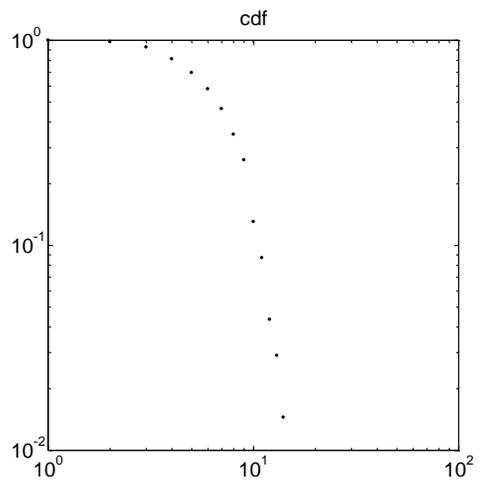
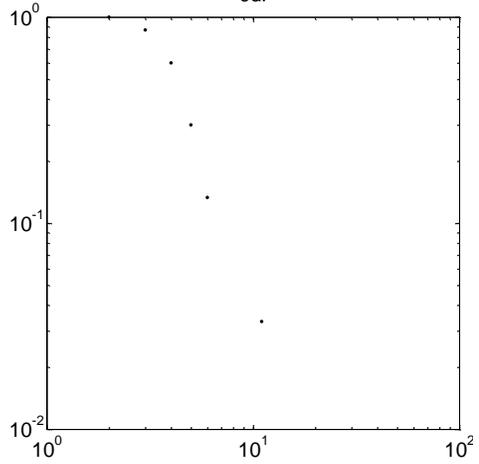
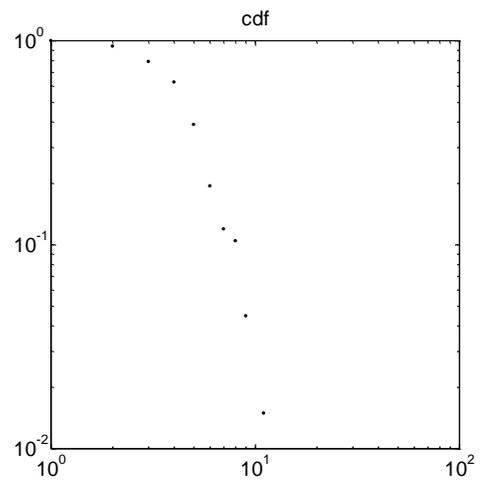

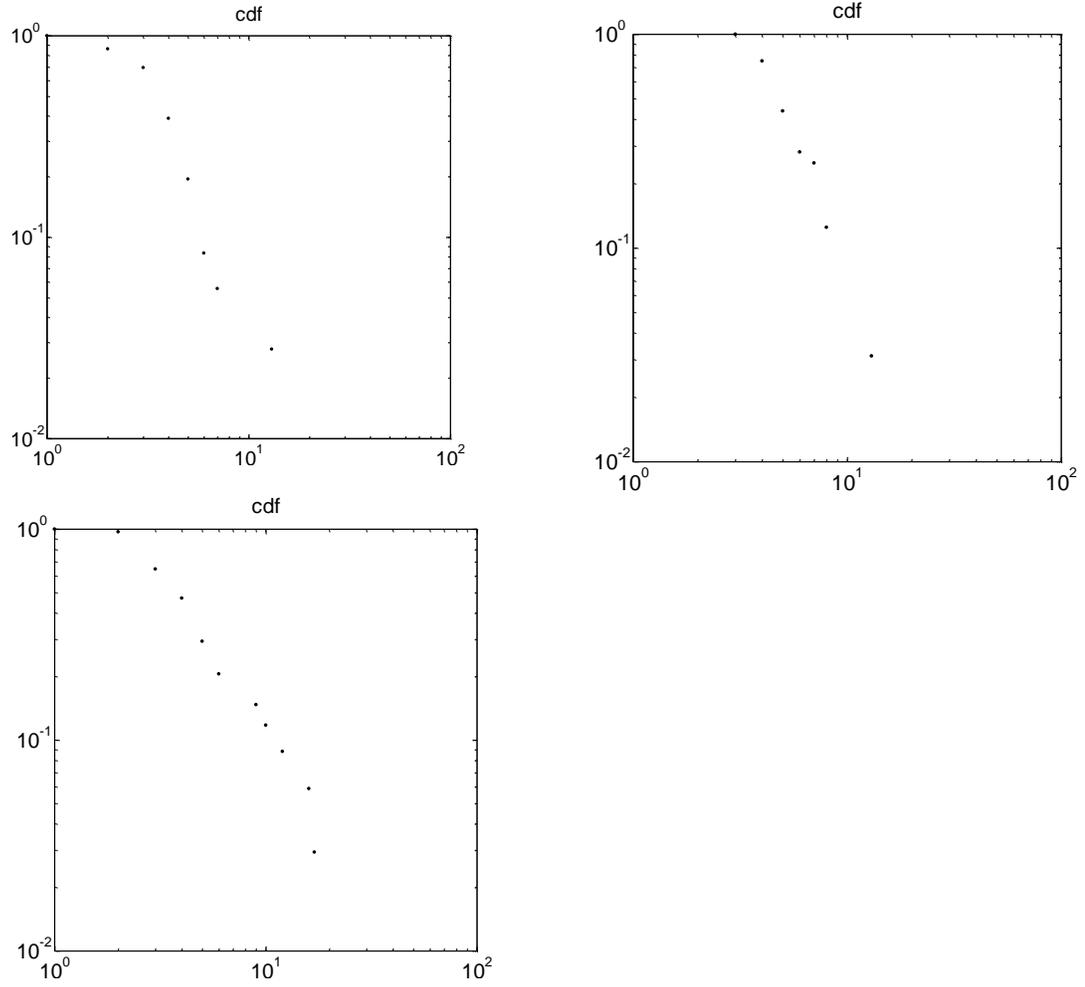

**Supplementary Figure S2 | Cumulative degree distribution of the 15 social networks studied.** BF23, BF70, BF71, ColoSpg, Corporate, Dolphins, Drugs, Galesburg, HighTech, HS, MathMethod, Prison, SawMill, Social3, Zackary.

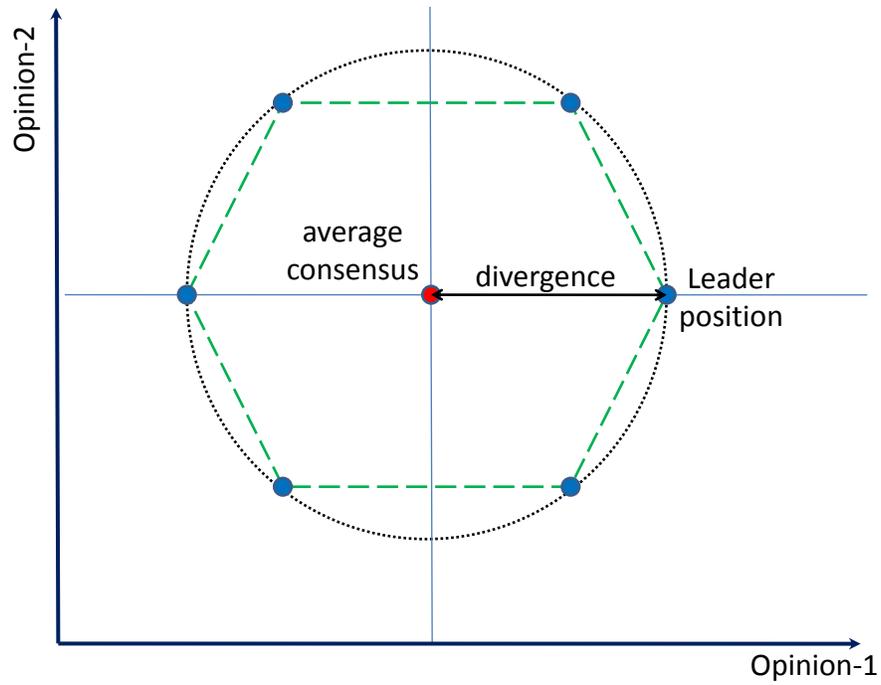

**Supplementary Figure S3 | Leaders with divergence.** Distribution of a leader's positions (blue points) around the average consensus value of the system (red point), which represents the centroid of the convex hull spanned by the leaders.

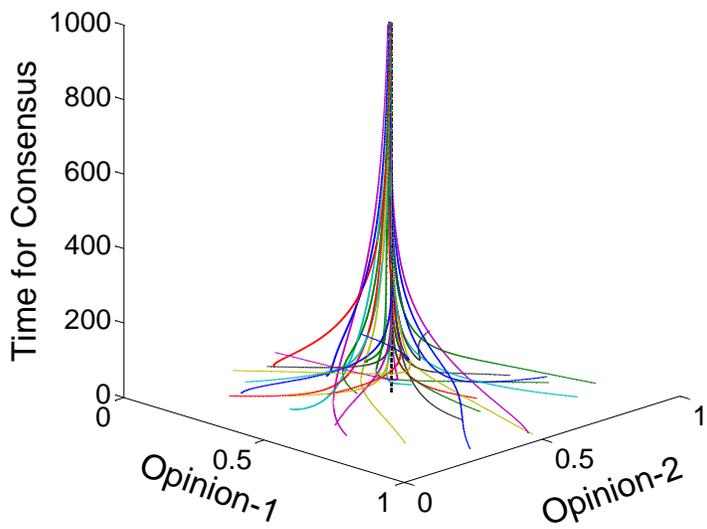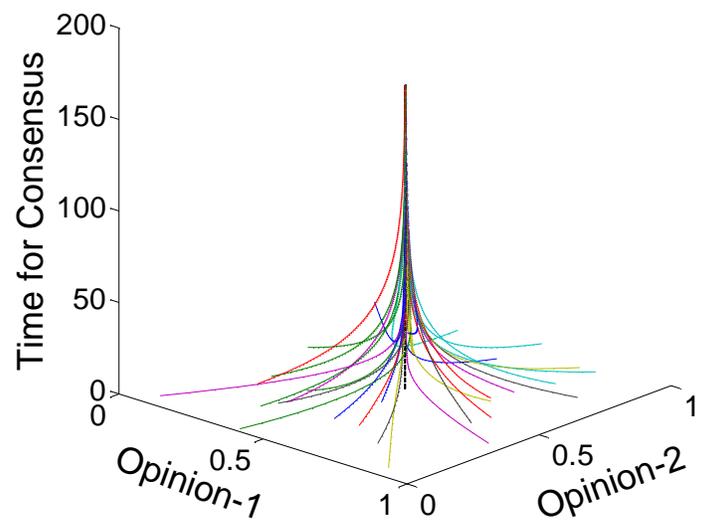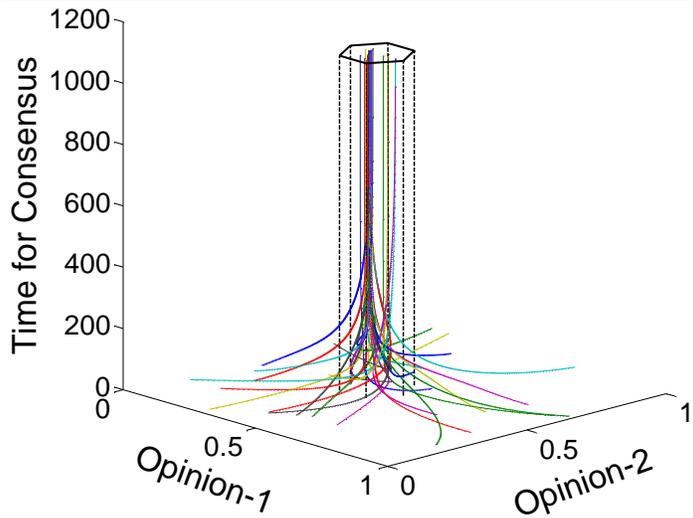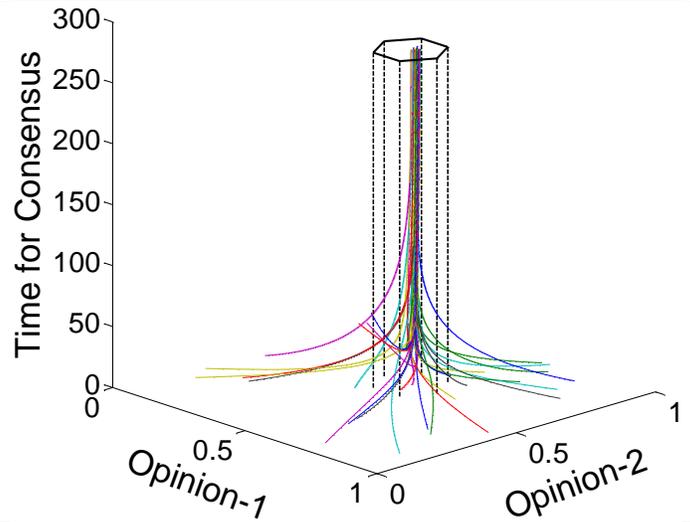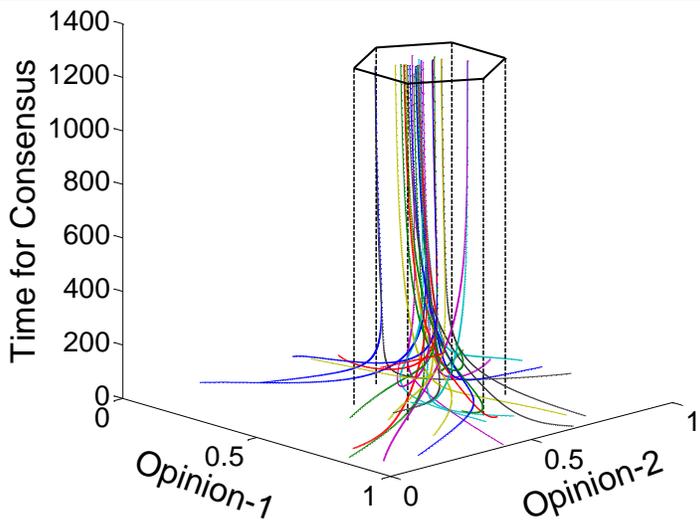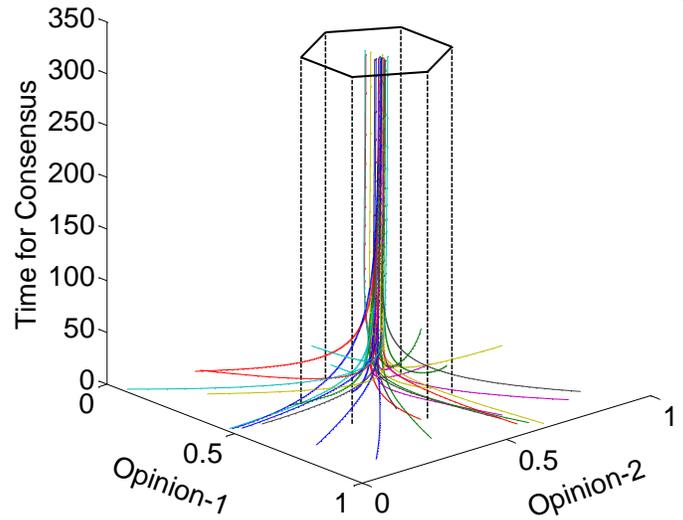

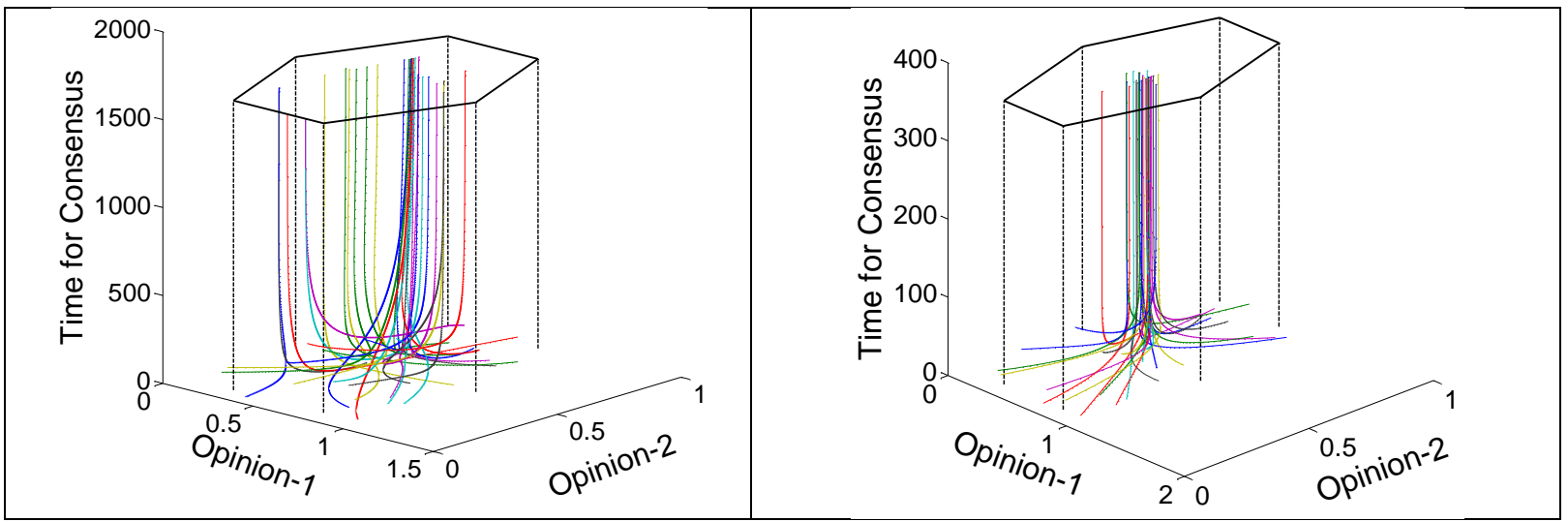

**Supplementary Figure S4 | Consensus for the Sawmill network**. Consensus with divergences (rows from top to bottom): 0.0, 0.1, 0.2, and 0.5, without PP (left column) and with PP (right column). The continuous lines indicate the states of the followers and the discontinuous lines indicate the states of the leaders during the consensus process.

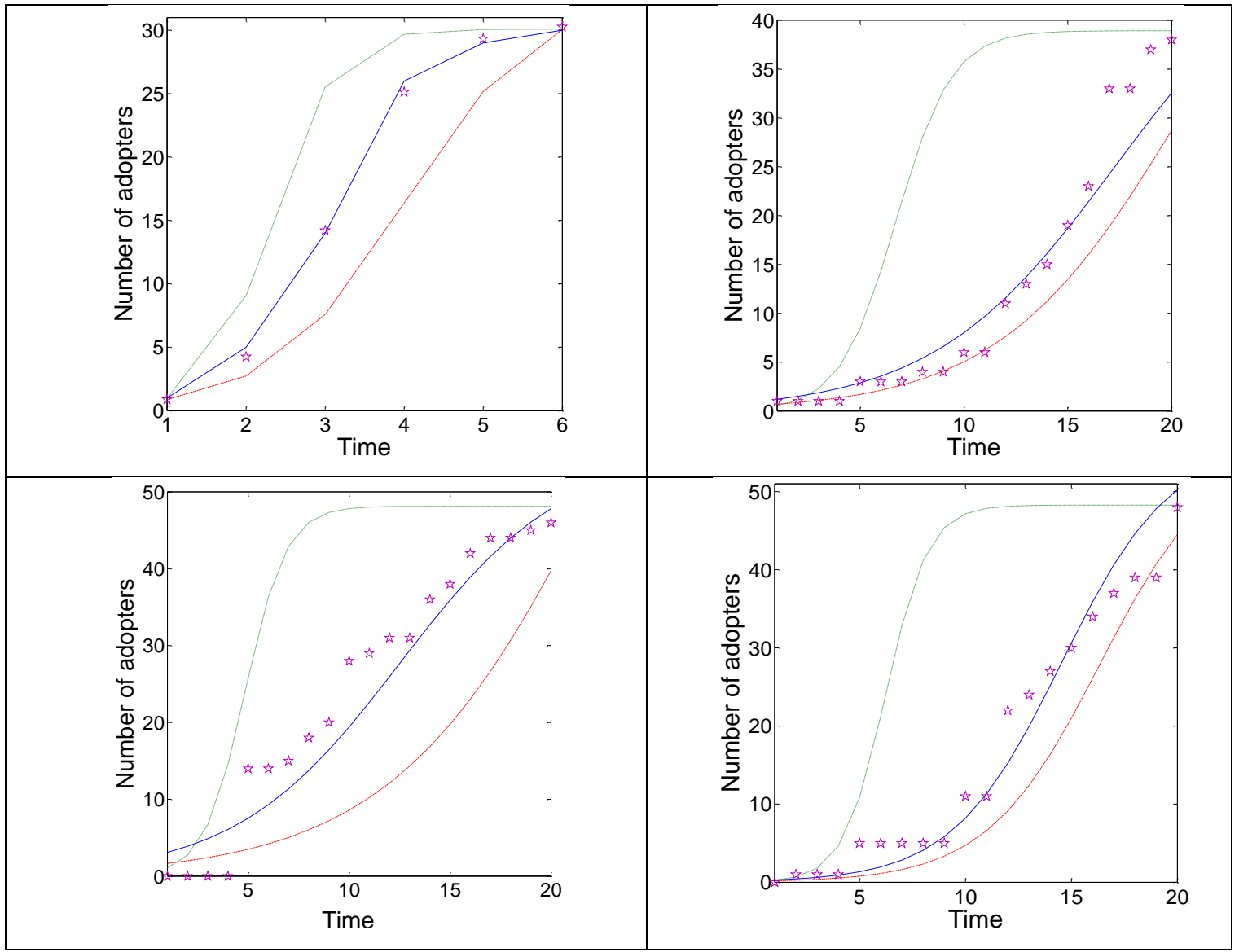

**Supplementary Figure S5 | Diffusion curves for empirical social networks.** The Mathematical Method (top left) and Brazilian Farmers Communities 23 (top right), 70 (bottom left), and 71(bottom right). The stars represent the empirical values of the cumulative number of adopters. The red lines represent diffusion without PP, the blue lines represent the diffusion process with moderate PP, and the green lines represent diffusion with strong PP.

**Supplementary Table S1**

Comparison of time for consensus for the Sawmill network with different values of divergence and the effect on time to reach consensus.

| Divergence | Average Time for Consensus | % Increase in Time for Consensus |
|---|---|---|
| 0 | 1,026.80 | - |
| 0.1 | 1,162.18 | 13.18 |
| 0.2 | 1,372.40 | 33.65 |
| 0.5 | 1,850.72 | 80.24 |

**Supplementary Table S2**

Normalized consensus time for the Sawmill network with no divergence in leaders' positions with respect to the system's average consensus value. All leaders' initial states were equal to the average consensus, i.e., the average of followers' initial states.

| | | \multicolumn{7}{c}{Random Emergence} | | | | | | |
|---|---|---|---|---|---|---|---|---|
| | **No PP** | **PL-decay** | | **Exp-decay** | | **Social** | | |
| | | α=-2 | α=-1.5 | β=-2 | β=-1.5 | δ=0.1 | δ=0.25 | δ=0.5 |
| | 1.00 | 0.22 | 0.15 | 0.88 | 0.72 | 0.49 | 0.20 | 0.06 |
| | | \multicolumn{7}{c}{Leader-Centrality Emergence} | | | | | | |
| | **No PP** | **PL-decay** | | **Exp-decay** | | **Social** | | |
| **Centrality** | | α=-2 | α=-1.5 | β=-2 | β=-1.5 | δ=0.1 | δ=0.25 | δ=0.5 |
| BC | 0.80 | 0.21 | 0.14 | 0.73 | 0.62 | 0.43 | 0.19 | 0.05 |
| CC | 0.84 | 0.21 | 0.14 | 0.77 | 0.65 | 0.45 | 0.19 | 0.05 |
| DC | 0.83 | 0.21 | 0.14 | 0.76 | 0.64 | 0.45 | 0.19 | 0.06 |
| EC | 0.96 | 0.23 | 0.15 | 0.87 | 0.73 | 0.50 | 0.20 | 0.06 |
| SC | 0.89 | 0.23 | 0.14 | 0.85 | 0.69 | 0.46 | 0.20 | 0.06 |

**Supplementary Table S3**

Normalized consensus time for the Sawmill network with divergence of 0.1 in leaders' positions with respect to the system's average consensus value, i.e., the length of the circumradius of the regular polygon spanned by the leaders is equal to 0.1.

|  | \multicolumn{7}{c}{Random Emergence} |
| --- | --- | --- | --- | --- | --- | --- | --- |
|  | No PP | PL-decay | | Exp-decay | | Social | | |
|  |  | α=-2 | α=-1.5 | β=-2 | β=-1.5 | δ=0.1 | δ=0.25 | δ=0.5 |
|  | 1.00 | 0.24 | 0.17 | 0.88 | 0.68 | 0.45 | 0.20 | 0.07 |
|  | \multicolumn{7}{c}{Leader-Centrality Emergence} |
|  | No PP | PL-decay | | Exp-decay | | Social | | |
| Centrality |  | α=-2 | α=-1.5 | β=-2 | β=-1.5 | δ=0.1 | δ=0.25 | δ=0.5 |
| BC | 0.72 | 0.19 | 0.13 | 0.65 | 0.54 | 0.38 | 0.17 | 0.05 |
| CC | 0.75 | 0.19 | 0.12 | 0.70 | 0.56 | 0.40 | 0.17 | 0.05 |
| DC | 0.74 | 0.19 | 0.12 | 0.71 | 0.59 | 0.40 | 0.17 | 0.05 |
| EC | 0.99 | 0.22 | 0.15 | 0.92 | 0.68 | 0.45 | 0.18 | 0.05 |
| SC | 0.82 | 0.20 | 0.13 | 0.76 | 0.62 | 0.41 | 0.18 | 0.05 |

**Supplementary Table S4**

Normalized consensus time for the Sawmill network with divergence of 0.2 in leaders' positions with respect to the system's average consensus value, i.e., the length of the circumradius of the regular polygon spanned by the leaders is equal to 0.2.

|  | \multicolumn{7}{c}{Random Emergence} |
| --- | --- | --- | --- | --- | --- | --- | --- |
|  | No PP | PL-decay | | Exp-decay | | Social | | |
|  |  | α=-2 | α=-1.5 | β=-2 | β=-1.5 | δ=0.1 | δ=0.25 | δ=0.5 |
|  | 0.96 | 0.23 | 0.16 | 0.87 | 0.66 | 0.43 | 0.19 | 0.06 |
|  | \multicolumn{7}{c}{Leader-Centrality Emergence} |
|  | No PP | PL-decay | | Exp-decay | | Social | | |
| Centrality |  | α=-2 | α=-1.5 | β=-2 | β=-1.5 | δ=0.1 | δ=0.25 | δ=0.5 |
| BC | 0.63 | 0.17 | 0.12 | 0.59 | 0.46 | 0.33 | 0.14 | 0.05 |
| CC | 0.66 | 0.16 | 0.12 | 0.58 | 0.49 | 0.33 | 0.14 | 0.04 |
| DC | 0.64 | 0.15 | 0.10 | 0.60 | 0.48 | 0.33 | 0.14 | 0.04 |
| EC | 1.00 | 0.19 | 0.13 | 0.85 | 0.63 | 0.37 | 0.15 | 0.05 |
| SC | 0.75 | 0.17 | 0.12 | 0.65 | 0.51 | 0.35 | 0.14 | 0.04 |

**Supplementary Table S5**

Normalized consensus time for the Sawmill network with divergence of 0.5 in leaders' positions with respect to the system's average consensus value, i.e., the length of the circumradius of the regular polygon spanned by the leaders is equal to 0.5.

|  | | Random Emergence | | | | | | |
|---|---|---|---|---|---|---|---|---|
|  | No PP | PL-decay | | Exp-decay | | Social | | |
|  |  | α=-2 | α=-1.5 | β=-2 | β=-1.5 | δ=0.1 | δ=0.25 | δ=0.5 |
|  | 1.00 | 0.23 | 0.16 | 0.87 | 0.67 | 0.40 | 0.17 | 0.06 |
|  | | Leader-Centrality Emergence | | | | | | |
|  | No PP | PL-decay | | Exp-decay | | Social | | |
| Centrality |  | α=-2 | α=-1.5 | β=-2 | β=-1.5 | δ=0.1 | δ=0.25 | δ=0.5 |
| BC | 0.59 | 0.15 | 0.11 | 0.53 | 0.43 | 0.27 | 0.11 | 0.04 |
| CC | 0.60 | 0.14 | 0.11 | 0.52 | 0.41 | 0.27 | 0.11 | 0.04 |
| DC | 0.60 | 0.12 | 0.08 | 0.54 | 0.43 | 0.27 | 0.10 | 0.03 |
| EC | 0.99 | 0.17 | 0.12 | 0.83 | 0.59 | 0.35 | 0.13 | 0.04 |
| SC | 0.72 | 0.14 | 0.10 | 0.60 | 0.47 | 0.29 | 0.11 | 0.04 |

**Supplementary Table S6**

Normalized consensus times for a random graph with 10 communities (10 leaders) with and without PP. Leaders emerged according to their global centrality values.

|  | | Random Emergence | | | | | | |
|---|---|---|---|---|---|---|---|---|
|  | No PP | PL-decay | | Exp-decay | | Social | | |
|  |  | α=-2 | α=-1.5 | β=-2 | β=-1.5 | δ=0.1 | δ=0.25 | δ=0.5 |
|  | 0.15 | 0.01 | 0.01 | 0.12 | 0.06 | 0.04 | 0.01 | 0.003 |
|  | | Leader-Centrality Emergence | | | | | | |
|  | No PP | PL-decay | | Exp-decay | | Social | | |
| Centrality |  | α=-2 | α=-1.5 | β=-2 | β=-1.5 | δ=0.1 | δ=0.25 | δ=0.5 |
| BC | 0.26 | 0.01 | 0.01 | 0.18 | 0.09 | 0.06 | 0.01 | 0.003 |
| CC | 0.46 | 0.01 | 0.01 | 0.32 | 0.16 | 0.10 | 0.02 | 0.002 |
| DC | 0.44 | 0.01 | 0.01 | 0.32 | 0.16 | 0.10 | 0.02 | 0.003 |
| EC | 1.00 | 0.01 | 0.01 | 0.62 | 0.27 | 0.16 | 0.03 | 0.003 |
| SC | 0.59 | 0.01 | 0.01 | 0.34 | 0.15 | 0.09 | 0.02 | 0.004 |

**Supplementary Table S7**

Normalized consensus times for a graph with 10 communities (10 leaders) with and without PP. Leaders emerged according to their centrality values by community.

|            |       | Random Emergence |              |              |              |              |              |              |
|------------|-------|------------------|--------------|--------------|--------------|--------------|--------------|--------------|
|            | No PP | PL-decay         |              | Exp-decay    |              | Social       |              |              |
|            |       | α=-2             | α=-1.5       | β=-2         | β=-1.5       | δ=0.1        | δ=0.25       | δ=0.5        |
|            | 1.00  | 0.06             | 0.04         | 0.77         | 0.41         | 0.27         | 0.08         | 0.02         |
|            |       | Leader-Centrality Emergence |    |              |              |              |              |              |
|            | No PP | PL-decay         |              | Exp-decay    |              | Social       |              |              |
| Centrality |       | α=2              | α=1.5        | β=2          | β=1.5        | δ=0.1        | δ=0.25       | δ=0.5        |
| BC         | 0.54  | 0.07             | 0.04         | 0.42         | 0.28         | 0.17         | 0.06         | 0.02         |
| CC         | 0.40  | 0.07             | 0.04         | 0.31         | 0.22         | 0.14         | 0.06         | 0.02         |
| DC         | 0.30  | 0.07             | 0.05         | 0.26         | 0.20         | 0.13         | 0.07         | 0.02         |
| EC         | 0.43  | 0.07             | 0.04         | 0.34         | 0.25         | 0.15         | 0.06         | 0.02         |
| SC         | 0.29  | 0.07             | 0.04         | 0.25         | 0.19         | 0.12         | 0.06         | 0.02         |

**Supplementary Table S8**

Cumulative average nodes in agreement for the Math Method network with and without PP and empirical values.

| Periods | Adopters (Empirical) | Avg. Adopters (Simulation) | | |
|---|---|---|---|---|
| | | No LR | α=-4 | α=-5 |
| 1 | 1  | 1.1  | 1.3  | 1.1  |
| 2 | 5  | 2.7  | 9.0  | 4.7  |
| 3 | 14 | 7.7  | 25.6 | 13.7 |
| 4 | 26 | 23.6 | 29.8 | 27.8 |
| 5 | 29 | 28.3 | 30   | 29.4 |
| 6 | 30 | 30   | 30   | 30   |

**Supplementary Table S9**

Cumulative average nodes in agreement for the Brazilian Farmers, Community 23 network with and without PP and empirical values.

| Periods | Adopters (Empirical) | Avg. Adopters (Simulation) | | |
|---|---|---|---|---|
| | | *No PP* | *α=-4* | *α=-5.9* |
| 1 | 1 | 0.8 | 1.0 | 0.8 |
| 2 | 1 | 1.3 | 1.8 | 1.2 |
| 3 | 1 | 1.2 | 2.0 | 1.1 |
| 4 | 1 | 1.5 | 3.0 | 1.5 |
| 5 | 3 | 1.8 | 7.2 | 1.7 |
| 6 | 3 | 2.0 | 15.8 | 2.2 |
| 7 | 3 | 2.3 | 23.4 | 2.4 |
| 8 | 4 | 2.5 | 26.0 | 4.1 |
| 9 | 4 | 4.0 | 32.3 | 9.0 |
| 10 | 6 | 5.6 | 35.9 | 10.9 |
| 11 | 6 | 7.2 | 37.2 | 11.5 |
| 12 | 11 | 8.9 | 38.5 | 12.5 |
| 13 | 13 | 9.5 | 38.9 | 13.5 |
| 14 | 15 | 10.3 | 38.9 | 15.8 |
| 15 | 19 | 12.3 | 38.9 | 17.3 |
| 16 | 23 | 13.4 | 38.9 | 19.8 |
| 17 | 33 | 19.0 | 38.9 | 24.6 |
| 18 | 33 | 24.9 | 38.9 | 26.9 |
| 19 | 37 | 26.4 | 38.9 | 30.4 |
| 20 | 38 | 27.1 | 38.9 | 32.9 |

**Supplementary Table S10**

Cumulative average nodes in agreement for the Brazilian Farmers, Community 70 network with and without PP and empirical values.

| Periods | Adopters (Empirical) | Avg. Adopters (Simulation) | | |
|---|---|---|---|---|
| | | *No PP* | *α=-4* | *α =-5.6* |
| 1 | 0 | 0.9 | 0.8 | 0.7 |
| 2 | 0 | 1.7 | 3.2 | 1.8 |
| 3 | 0 | 1.8 | 9.5 | 3.0 |
| 4 | 0 | 3.3 | 14.3 | 4.5 |
| 5 | 14 | 4.9 | 23.4 | 5.6 |
| 6 | 14 | 4.7 | 36.2 | 8.6 |
| 7 | 15 | 5.8 | 45.4 | 13.7 |
| 8 | 18 | 6.4 | 47.7 | 16.5 |
| 9 | 20 | 6.6 | 47.8 | 18.9 |
| 10 | 28 | 8.3 | 47.8 | 21.0 |
| 11 | 29 | 10.2 | 47.8 | 23.5 |
| 12 | 31 | 11.8 | 47.8 | 25.9 |
| 13 | 31 | 14.0 | 47.8 | 28.4 |
| 14 | 36 | 16.5 | 47.8 | 30.7 |
| 15 | 38 | 19.1 | 47.8 | 33.1 |
| 16 | 42 | 22.7 | 47.8 | 37.8 |
| 17 | 44 | 27.3 | 47.8 | 42.1 |
| 18 | 44 | 31.7 | 47.8 | 45.6 |
| 19 | 45 | 35.4 | 47.8 | 47.4 |
| 20 | 46 | 39.0 | 47.8 | 47.5 |

**Supplementary Table S11**

Cumulative average nodes in agreement for the Brazilian Farmers, Community 71 network with and without PP and empirical values.

| Periods | Adopters (Empirical) | Avg. Adopters (Simulation) | | |
|---|---|---|---|---|
| | | *No PP* | *α=-4* | *α=-6.6* |
| 1 | 0 | 0.5 | 0.7 | 0.7 |
| 2 | 1 | 0.9 | 1.9 | 1.1 |
| 3 | 1 | 1.2 | 4.0 | 1.4 |
| 4 | 1 | 1.1 | 7.3 | 1.6 |
| 5 | 5 | 1.0 | 11.8 | 3.2 |
| 6 | 5 | 1.2 | 21.6 | 5.5 |
| 7 | 5 | 2.1 | 32.2 | 6.4 |
| 8 | 5 | 2.9 | 37.3 | 6.6 |
| 9 | 5 | 3.6 | 46.0 | 7.0 |
| 10 | 11 | 4.6 | 47.8 | 7.9 |
| 11 | 11 | 6.1 | 48.4 | 9.6 |
| 12 | 22 | 8.5 | 48.4 | 12.7 |
| 13 | 24 | 11.7 | 48.4 | 17.3 |
| 14 | 27 | 16.4 | 48.4 | 23.8 |
| 15 | 30 | 21.9 | 48.4 | 31.5 |
| 16 | 34 | 26.6 | 48.4 | 38.2 |
| 17 | 37 | 30.8 | 48.4 | 43.1 |
| 18 | 39 | 36.2 | 48.4 | 45.9 |
| 19 | 39 | 40.8 | 48.4 | 47.4 |
| 20 | 48 | 44.5 | 48.4 | 47.6 |

**Supplementary Table S12**

Normalized consensus times for the BA random graph with divergence of 0.1.

|  | Random Emergence | | | | | | | |
|---|---|---|---|---|---|---|---|---|
|  | **No PP** | **PL-decay** | | **Exp-decay** | | **Social** | | |
|  |  | α=-2 | α=-1.5 | β=-2 | β=-1.5 | δ=0.1 | δ=0.25 | δ=0.5 |
|  | 1.00 | 0.44 | 0.31 | 0.75 | 0.67 | 0.84 | 0.38 | 0.13 |
|  | Leader-Centrality Emergence | | | | | | | |
|  | **No PP** | **PL-decay** | | **Exp-decay** | | **Social** | | |
| **Centrality** |  | α=-2 | α=-1.5 | β=-2 | β=-1.5 | δ=0.1 | δ=0.25 | δ=0.5 |
| BC | 0.89 | 0.33 | 0.26 | 0.73 | 0.61 | 0.41 | 0.23 | 0.09 |
| CC | 0.91 | 0.34 | 0.27 | 0.78 | 0.65 | 0.42 | 0.22 | 0.08 |
| DC | 0.89 | 0.33 | 0.26 | 0.75 | 0.62 | 0.41 | 0.23 | 0.09 |
| EC | 0.93 | 0.34 | 0.27 | 0.77 | 0.64 | 0.42 | 0.22 | 0.08 |
| SC | 0.90 | 0.34 | 0.27 | 0.76 | 0.63 | 0.41 | 0.22 | 0.08 |

**Supplementary Table S13**

Normalized consensus times for the ER random graph with divergence of 0.1.

|  | Random Emergence | | | | | | | |
|---|---|---|---|---|---|---|---|---|
|  | **No PP** | **PL-decay** | | **Exp-decay** | | **Social** | | |
|  |  | α=-2 | α=-1.5 | β=-2 | β=-1.5 | δ=0.1 | δ=0.25 | δ=0.5 |
|  | 1.00 | 0.34 | 0.23 | 0.90 | 0.75 | 0.61 | 0.31 | 0.10 |
|  | Leader-Centrality Emergence | | | | | | | |
|  | **No PP** | **PL-decay** | | **Exp-decay** | | **Social** | | |
| **Centrality** |  | α=-2 | α=-1.5 | β=-2 | β=-1.5 | δ=0.1 | δ=0.25 | δ=0.5 |
| BC | 0.81 | 0.10 | 0.06 | 0.67 | 0.47 | 0.28 | 0.10 | 0.05 |
| CC | 0.81 | 0.10 | 0.06 | 0.67 | 0.46 | 0.27 | 0.09 | 0.02 |
| DC | 0.81 | 0.10 | 0.06 | 0.69 | 0.48 | 0.28 | 0.10 | 0.03 |
| EC | 0.82 | 0.13 | 0.08 | 0.69 | 0.48 | 0.30 | 0.11 | 0.03 |
| SC | 0.82 | 0.13 | 0.08 | 0.68 | 0.48 | 0.31 | 0.11 | 0.02 |

**Supplementary Table S14**

Normalized consensus times for the Corporate network with divergence of 0.1.

|  | Random Emergence | | | | | | | |
|---|---|---|---|---|---|---|---|---|
|  | No PP | PL-decay | | Exp-decay | | Social | | |
|  |  | α=-2 | α=-1.5 | β=-2 | β=-1.5 | δ=0.1 | δ=0.25 | δ=0.5 |
|  | 1.00 | 0.36 | 0.22 | 0.96 | 0.92 | 0.94 | 0.37 | 0.11 |
|  | Leader-Centrality Emergence | | | | | | | |
|  | No PP | PL-decay | | Exp-decay | | Social | | |
| Centrality |  | α=-2 | α=-1.5 | β=-2 | β=-1.5 | δ=0.1 | δ=0.25 | δ=0.5 |
| BC | 0.67 | 0.02 | 0.01 | 0.52 | 0.35 | 0.27 | 0.08 | 0.01 |
| CC | 0.70 | 0.03 | 0.01 | 0.54 | 0.35 | 0.28 | 0.09 | 0.01 |
| DC | 0.69 | 0.05 | 0.04 | 0.52 | 0.36 | 0.29 | 0.10 | 0.02 |
| EC | 0.67 | 0.02 | 0.01 | 0.52 | 0.36 | 0.27 | 0.08 | 0.01 |
| SC | 0.68 | 0.02 | 0.01 | 0.51 | 0.35 | 0.27 | 0.08 | 0.01 |

**Supplementary Table S15**

Normalized consensus times for the Drugs network with divergence of 0.1.

|  | Random Emergence | | | | | | | |
|---|---|---|---|---|---|---|---|---|
|  | No PP | PL-decay | | Exp-decay | | Social | | |
|  |  | α=-2 | α=-1.5 | β=-2 | β=-1.5 | δ=0.1 | δ=0.25 | δ=0.5 |
|  | 1.00 | 0.17 | 0.12 | 0.89 | 0.64 | 0.53 | 0.24 | 0.08 |
|  | Leader-Centrality Emergence | | | | | | | |
|  | No PP | PL-decay | | Exp-decay | | Social | | |
| Centrality |  | α=-2 | α=-1.5 | β=-2 | β=-1.5 | δ=0.1 | δ=0.25 | δ=0.5 |
| BC | 0.97 | 0.07 | 0.04 | 0.79 | 0.51 | 0.40 | 0.12 | 0.02 |
| CC | 0.79 | 0.08 | 0.05 | 0.73 | 0.50 | 0.40 | 0.12 | 0.02 |
| DC | 0.77 | 0.04 | 0.06 | 0.66 | 0.45 | 0.36 | 0.13 | 0.06 |
| EC | 0.80 | 0.11 | 0.08 | 0.71 | 0.53 | 0.43 | 0.13 | 0.04 |
| SC | 0.80 | 0.11 | 0.08 | 0.72 | 0.50 | 0.42 | 0.13 | 0.04 |

**Supplementary Table S16**

Normalized consensus times for the Prison network with divergence of 0.1.

| | | Random Emergence | | | | | | |
|---|---|---|---|---|---|---|---|---|
| | **No PP** | **PL-decay** | | **Exp-decay** | | **Social** | | |
| | | $\alpha=-2$ | $\alpha=-1.5$ | $\beta=-2$ | $\beta=-1.5$ | $\delta=0.1$ | $\delta=0.25$ | $\delta=0.5$ |
| | 1.00 | 0.26 | 0.19 | 0.93 | 0.70 | 0.49 | 0.23 | 0.08 |
| | | Leader-Centrality Emergence | | | | | | |
| | **No PP** | **PL-decay** | | **Exp-decay** | | **Social** | | |
| **Centrality** | | $\alpha=-2$ | $\alpha=-1.5$ | $\beta=-2$ | $\beta=-1.5$ | $\delta=0.1$ | $\delta=0.25$ | $\delta=0.5$ |
| BC | 0.93 | 0.17 | 0.11 | 0.85 | 0.68 | 0.45 | 0.17 | 0.04 |
| CC | 0.97 | 0.17 | 0.11 | 0.86 | 0.66 | 0.45 | 0.17 | 0.04 |
| DC | 0.95 | 0.17 | 0.10 | 0.85 | 0.65 | 0.44 | 0.17 | 0.04 |
| EC | 0.98 | 0.17 | 0.11 | 0.84 | 0.68 | 0.44 | 0.17 | 0.04 |
| SC | 0.94 | 0.17 | 0.10 | 0.85 | 0.65 | 0.43 | 0.17 | 0.04 |

**Supplementary Table S17**

Normalized consensus times for the Zachary network with divergence of 0.1.

| | | Random Emergence | | | | | | |
|---|---|---|---|---|---|---|---|---|
| | **No PP** | **PL-decay** | | **Exp-decay** | | **Social** | | |
| | | $\alpha=-2$ | $\alpha=-1.5$ | $\beta=-2$ | $\beta=-1.5$ | $\delta=0.1$ | $\delta=0.25$ | $\delta=0.5$ |
| | 1.00 | 0.27 | 0.19 | 0.94 | 0.67 | 0.39 | 0.19 | 0.08 |
| | | Leader-Centrality Emergence | | | | | | |
| | **No PP** | **PL-decay** | | **Exp-decay** | | **Social** | | |
| **Centrality** | | $\alpha=-2$ | $\alpha=-1.5$ | $\beta=-2$ | $\beta=-1.5$ | $\delta=0.1$ | $\delta=0.25$ | $\delta=0.5$ |
| BC | 0.73 | 0.22 | 0.17 | 0.63 | 0.51 | 0.32 | 0.16 | 0.06 |
| CC | 0.75 | 0.22 | 0.16 | 0.66 | 0.50 | 0.32 | 0.15 | 0.06 |
| DC | 0.76 | 0.22 | 0.17 | 0.64 | 0.51 | 0.32 | 0.16 | 0.06 |
| EC | 0.75 | 0.21 | 0.17 | 0.64 | 0.50 | 0.33 | 0.16 | 0.05 |
| SC | 0.76 | 0.22 | 0.17 | 0.65 | 0.49 | 0.31 | 0.16 | 0.06 |

**Supplementary Table S18**

Normalized consensus times for the Colorado Springs network with divergence of 0.1.

|  | Random Emergence | | | | | | | |
|---|---|---|---|---|---|---|---|---|
|  | No PP | PL-decay | | Exp-decay | | Social | | |
|  |  | α=-2 | α=-1.5 | β=-2 | β=-1.5 | δ=0.1 | δ=0.25 | δ=0.5 |
|  | 0.83 | 0.22 | 0.15 | 0.76 | 0.62 | 0.56 | 0.37 | 0.13 |
|  | Leader-Centrality Emergence | | | | | | | |
|  | No PP | PL-decay | | Exp-decay | | Social | | |
| Centrality |  | α=-2 | α=-1.5 | β=-2 | β=-1.5 | δ=0.1 | δ=0.25 | δ=0.5 |
| BC | 1.00 | 0.11 | 0.07 | 0.87 | 0.69 | 0.53 | 0.25 | 0.04 |
| CC | 0.90 | 0.07 | 0.03 | 0.82 | 0.53 | 0.27 | 0.27 | 0.04 |
| DC | 0.89 | 0.13 | 0.09 | 0.82 | 0.65 | 0.50 | 0.24 | 0.06 |
| EC | 0.85 | 0.13 | 0.06 | 0.75 | 0.62 | 0.47 | 0.22 | 0.04 |
| SC | 0.95 | 0.17 | 0.12 | 0.86 | 0.69 | 0.54 | 0.28 | 0.07 |

**Supplementary Table S19**

Normalized consensus times for the Dolphins network with divergence of 0.1.

|  | Random Emergence | | | | | | | |
|---|---|---|---|---|---|---|---|---|
|  | No PP | PL-decay | | Exp-decay | | Social | | |
|  |  | α=-2 | α=-1.5 | β=-2 | β=-1.5 | δ=0.1 | δ=0.25 | δ=0.5 |
|  | 0.82 | 0.25 | 0.16 | 0.76 | 0.61 | 0.38 | 0.20 | 0.08 |
|  | Leader-Centrality Emergence | | | | | | | |
|  | No PP | PL-decay | | Exp-decay | | Social | | |
| Centrality |  | α=-2 | α=-1.5 | β=-2 | β=-1.5 | δ=0.1 | δ=0.25 | δ=0.5 |
| BC | 0.77 | 0.17 | 0.11 | 0.70 | 0.56 | 0.37 | 0.15 | 0.06 |
| CC | 0.92 | 0.17 | 0.12 | 0.65 | 0.64 | 0.43 | 0.18 | 0.05 |
| DC | 0.72 | 0.18 | 0.12 | 0.64 | 0.57 | 0.39 | 0.17 | 0.06 |
| EC | 0.96 | 0.21 | 0.14 | 0.99 | 0.79 | 0.52 | 0.21 | 0.06 |
| SC | 1.00 | 0.22 | 0.14 | 0.97 | 0.77 | 0.54 | 0.21 | 0.06 |

**Supplementary Table S20**

Normalized consensus times for the Galesburg network with divergence of 0.1.

| | | Random Emergence | | | | | | |
|---|---|---|---|---|---|---|---|---|
| | No PP | PL-decay | | Exp-decay | | Social | | |
| | | $\alpha$=-2 | $\alpha$=-1.5 | $\beta$=-2 | $\beta$=-1.5 | $\delta$=0.1 | $\delta$=0.25 | $\delta$=0.5 |
| | 0.83 | 0.23 | 0.16 | 0.71 | 0.59 | 0.39 | 0.17 | 0.06 |
| | | Leader-Centrality Emergence | | | | | | |
| | No PP | PL-decay | | Exp-decay | | Social | | |
| Centrality | | $\alpha$=-2 | $\alpha$=-1.5 | $\beta$=-2 | $\beta$=-1.5 | $\delta$=0.1 | $\delta$=0.25 | $\delta$=0.5 |
| BC | 0.81 | 0.21 | 0.14 | 0.73 | 0.59 | 0.37 | 0.16 | 0.05 |
| CC | 0.98 | 0.24 | 0.16 | 0.87 | 0.66 | 0.40 | 0.16 | 0.05 |
| DC | 0.98 | 0.23 | 0.16 | 0.84 | 0.64 | 0.41 | 0.17 | 0.05 |
| EC | 1.00 | 0.23 | 0.17 | 0.86 | 0.65 | 0.41 | 0.17 | 0.06 |
| SC | 1.00 | 0.24 | 0.17 | 0.87 | 0.68 | 0.41 | 0.17 | 0.06 |

**Supplementary Table S21**

Normalized consensus times for the HS network with divergence of 0.1.

| | | Random Emergence | | | | | | |
|---|---|---|---|---|---|---|---|---|
| | No PP | PL-decay | | Exp-decay | | Social | | |
| | | $\alpha$=-2 | $\alpha$=-1.5 | $\beta$=-2 | $\beta$=-1.5 | $\delta$=0.1 | $\delta$=0.25 | $\delta$=0.5 |
| | 0.75 | 0.20 | 0.16 | 0.69 | 0.51 | 0.35 | 0.18 | 0.07 |
| | | Leader-Centrality Emergence | | | | | | |
| | No PP | PL-decay | | Exp-decay | | Social | | |
| Centrality | | $\alpha$=-2 | $\alpha$=-1.5 | $\beta$=-2 | $\beta$=-1.5 | $\delta$=0.1 | $\delta$=0.25 | $\delta$=0.5 |
| BC | 0.65 | 0.18 | 0.14 | 0.56 | 0.43 | 0.28 | 0.15 | 0.06 |
| CC | 0.99 | 0.20 | 0.13 | 0.81 | 0.60 | 0.35 | 0.14 | 0.06 |
| DC | 0.69 | 0.17 | 0.11 | 0.59 | 0.46 | 0.27 | 0.13 | 0.07 |
| EC | 0.99 | 0.21 | 0.15 | 0.81 | 0.57 | 0.34 | 0.14 | 0.04 |
| SC | 1.00 | 0.21 | 0.15 | 0.78 | 0.59 | 0.34 | 0.14 | 0.04 |

**Supplementary Table S22**

Normalized consensus times for the High Tech network with divergence of 0.1.

| | | Random Emergence | | | | | | |
|---|---|---|---|---|---|---|---|---|
| | **No PP** | **PL-decay** | | **Exp-decay** | | **Social** | | |
| | | α=-2 | α=-1.5 | β=-2 | β=-1.5 | δ=0.1 | δ=0.25 | δ=0.5 |
| | 0.93 | 0.25 | 0.19 | 0.83 | 0.67 | 0.41 | 0.18 | 0.07 |
| | | Leader-Centrality Emergence | | | | | | |
| | **No PP** | **PL-decay** | | **Exp-decay** | | **Social** | | |
| **Centrality** | | α=-2 | α=-1.5 | β=-2 | β=-1.5 | δ=0.1 | δ=0.25 | δ=0.5 |
| BC | 0.78 | 0.22 | 0.16 | 0.70 | 0.56 | 0.37 | 0.16 | 0.06 |
| CC | 0.97 | 0.24 | 0.17 | 0.84 | 0.66 | 0.41 | 0.18 | 0.06 |
| DC | 0.93 | 0.24 | 0.18 | 0.83 | 0.67 | 0.40 | 0.18 | 0.06 |
| EC | 0.98 | 0.25 | 0.19 | 0.88 | 0.68 | 0.42 | 0.19 | 0.07 |
| SC | 1.00 | 0.27 | 0.19 | 0.88 | 0.68 | 0.44 | 0.19 | 0.06 |

**Supplementary Table S23**

Normalized consensus times for the Math Method network with divergence of 0.1.

| | | Random Emergence | | | | | | |
|---|---|---|---|---|---|---|---|---|
| | **No PP** | **PL-decay** | | **Exp-decay** | | **Social** | | |
| | | α=-2 | α=-1.5 | β=-2 | β=-1.5 | δ=0.1 | δ=0.25 | δ=0.5 |
| | 0.87 | 0.26 | 0.19 | 0.77 | 0.61 | 0.41 | 0.19 | 0.07 |
| | | Leader-Centrality Emergence | | | | | | |
| | **No PP** | **PL-decay** | | **Exp-decay** | | **Social** | | |
| **Centrality** | | α=-2 | α=-1.5 | β=-2 | β=-1.5 | δ=0.1 | δ=0.25 | δ=0.5 |
| BC | 0.63 | 0.22 | 0.16 | 0.59 | 0.50 | 0.35 | 0.17 | 0.06 |
| CC | 1.00 | 0.26 | 0.19 | 0.87 | 0.71 | 0.47 | 0.19 | 0.07 |
| DC | 0.98 | 0.26 | 0.18 | 0.86 | 0.70 | 0.46 | 0.19 | 0.07 |
| EC | 0.96 | 0.27 | 0.18 | 0.90 | 0.74 | 0.45 | 0.20 | 0.07 |
| SC | 1.00 | 0.26 | 0.19 | 0.86 | 0.71 | 0.44 | 0.20 | 0.07 |

**Supplementary Table S24**

Normalized consensus times for the Sawmill network with divergence of 0.1.

| | | Random Emergence | | | | | | |
|---|---|---|---|---|---|---|---|---|
| | No PP | PL-decay | | Exp-decay | | Social | | |
| | | α=-2 | α=-1.5 | β=-2 | β=-1.5 | δ=0.1 | δ=0.25 | δ=0.5 |
| | 1.00 | 0.24 | 0.17 | 0.88 | 0.68 | 0.45 | 0.20 | 0.07 |
| | | Leader-Centrality Emergence | | | | | | |
| | No PP | PL-decay | | Exp-decay | | Social | | |
| Centrality | | α=-2 | α=-1.5 | β=-2 | β=-1.5 | δ=0.1 | δ=0.25 | δ=0.5 |
| BC | 0.72 | 0.19 | 0.13 | 0.65 | 0.54 | 0.38 | 0.17 | 0.05 |
| CC | 0.75 | 0.19 | 0.12 | 0.70 | 0.56 | 0.40 | 0.17 | 0.05 |
| DC | 0.74 | 0.19 | 0.12 | 0.71 | 0.59 | 0.40 | 0.17 | 0.05 |
| EC | 0.99 | 0.22 | 0.15 | 0.92 | 0.68 | 0.45 | 0.18 | 0.05 |
| SC | 0.82 | 0.20 | 0.13 | 0.76 | 0.62 | 0.41 | 0.18 | 0.05 |

**Supplementary Table S25**

Normalized consensus times for the Social3 network with divergence of 0.1.

| | | Random Emergence | | | | | | |
|---|---|---|---|---|---|---|---|---|
| | No PP | PL-decay | | Exp-decay | | Social | | |
| | | α=-2 | α=-1.5 | β=-2 | β=-1.5 | δ=0.1 | δ=0.25 | δ=0.5 |
| | 0.92 | 0.33 | 0.24 | 0.81 | 0.67 | 0.46 | 0.24 | 0.11 |
| | | Leader-Centrality Emergence | | | | | | |
| | No PP | PL-decay | | Exp-decay | | Social | | |
| Centrality | | α=-2 | α=-1.5 | β=-2 | β=-1.5 | δ=0.1 | δ=0.25 | δ=0.5 |
| BC | 0.92 | 0.31 | 0.24 | 0.79 | 0.68 | 0.42 | 0.21 | 0.07 |
| CC | 0.98 | 0.32 | 0.23 | 0.84 | 0.70 | 0.45 | 0.22 | 0.08 |
| DC | 1.00 | 0.33 | 0.25 | 0.87 | 0.70 | 0.45 | 0.23 | 0.08 |
| EC | 0.96 | 0.32 | 0.24 | 0.88 | 0.74 | 0.46 | 0.23 | 0.08 |
| SC | 0.92 | 0.32 | 0.25 | 0.86 | 0.69 | 0.46 | 0.22 | 0.09 |

**Supplementary Table S26**

Normalized consensus times for the BA random graph with divergence of 0.2.

| | | Random Emergence | | | | | | |
|---|---|---|---|---|---|---|---|---|
| | **No PP** | **PL-decay** | | **Exp-decay** | | **Social** | | |
| | | α=-2 | α=-1.5 | β=-2 | β=-1.5 | δ=0.1 | δ=0.25 | δ=0.5 |
| | 1.00 | 0.42 | 0.29 | 0.94 | 0.85 | 0.79 | 0.33 | 0.11 |
| | | Leader-Centrality Emergence | | | | | | |
| | **No PP** | **PL-decay** | | **Exp-decay** | | **Social** | | |
| **Centrality** | | α=-2 | α=-1.5 | β=-2 | β=-1.5 | δ=0.1 | δ=0.25 | δ=0.5 |
| BC | 0.69 | 0.28 | 0.22 | 0.60 | 0.51 | 0.34 | 0.18 | 0.08 |
| CC | 0.74 | 0.29 | 0.22 | 0.64 | 0.54 | 0.35 | 0.18 | 0.08 |
| DC | 0.70 | 0.28 | 0.22 | 0.60 | 0.50 | 0.33 | 0.19 | 0.08 |
| EC | 0.72 | 0.28 | 0.23 | 0.63 | 0.53 | 0.34 | 0.18 | 0.07 |
| SC | 0.71 | 0.28 | 0.22 | 0.62 | 0.53 | 0.34 | 0.18 | 0.08 |

**Supplementary Table S27**

Normalized consensus times for the ER random graph with divergence of 0.2.

| | | Random Emergence | | | | | | |
|---|---|---|---|---|---|---|---|---|
| | **No PP** | **PL-decay** | | **Exp-decay** | | **Social** | | |
| | | α=-2 | α=-1.5 | β=-2 | β=-1.5 | δ=0.1 | δ=0.25 | δ=0.5 |
| | 1.00 | 0.27 | 0.18 | 0.88 | 0.70 | 0.45 | 0.20 | 0.07 |
| | | Leader-Centrality Emergence | | | | | | |
| | **No PP** | **PL-decay** | | **Exp-decay** | | **Social** | | |
| **Centrality** | | α=-2 | α=-1.5 | β=-2 | β=-1.5 | δ=0.1 | δ=0.25 | δ=0.5 |
| BC | 0.67 | 0.21 | 0.16 | 0.57 | 0.46 | 0.33 | 0.18 | 0.07 |
| CC | 0.78 | 0.22 | 0.14 | 0.67 | 0.54 | 0.32 | 0.14 | 0.03 |
| DC | 0.70 | 0.22 | 0.18 | 0.58 | 0.45 | 0.36 | 0.20 | 0.08 |
| EC | 0.74 | 0.20 | 0.14 | 0.63 | 0.50 | 0.32 | 0.15 | 0.05 |
| SC | 0.75 | 0.23 | 0.17 | 0.64 | 0.51 | 0.36 | 0.18 | 0.07 |

**Supplementary Table S28**

Normalized consensus times for the Corporate network with divergence of 0.2.

| | | Random Emergence | | | | | | |
|---|---|---|---|---|---|---|---|---|
| | No PP | PL-decay | | Exp-decay | | Social | | |
| | | $\alpha=-2$ | $\alpha=-1.5$ | $\beta=-2$ | $\beta=-1.5$ | $\delta=0.1$ | $\delta=0.25$ | $\delta=0.5$ |
| | 1.00 | 0.19 | 0.11 | 0.78 | 0.64 | 0.45 | 0.17 | 0.04 |
| | | Leader-Centrality Emergence | | | | | | |
| | No PP | PL-decay | | Exp-decay | | Social | | |
| Centrality | | $\alpha=-2$ | $\alpha=-1.5$ | $\beta=-2$ | $\beta=-1.5$ | $\delta=0.1$ | $\delta=0.25$ | $\delta=0.5$ |
| BC | 0.22 | 0.01 | 0.00 | 0.17 | 0.11 | 0.09 | 0.03 | 0.002 |
| CC | 0.28 | 0.02 | 0.02 | 0.23 | 0.15 | 0.11 | 0.03 | 0.002 |
| DC | 0.24 | 0.04 | 0.03 | 0.20 | 0.13 | 0.12 | 0.04 | 0.007 |
| EC | 0.22 | 0.01 | 0.00 | 0.18 | 0.11 | 0.09 | 0.03 | 0.002 |
| SC | 0.23 | 0.01 | 0.00 | 0.18 | 0.11 | 0.09 | 0.03 | 0.002 |

**Supplementary Table S29**

Normalized consensus times for the Drugs network with divergence of 0.2.

| | | Random Emergence | | | | | | |
|---|---|---|---|---|---|---|---|---|
| | No PP | PL-decay | | Exp-decay | | Social | | |
| | | $\alpha=-2$ | $\alpha=-1.5$ | $\beta=-2$ | $\beta=-1.5$ | $\delta=0.1$ | $\delta=0.25$ | $\delta=0.5$ |
| | 1.00 | 0.15 | 0.09 | 0.76 | 0.45 | 0.44 | 0.19 | 0.05 |
| | | Leader-Centrality Emergence | | | | | | |
| | No PP | PL-decay | | Exp-decay | | Social | | |
| Centrality | | $\alpha=-2$ | $\alpha=-1.5$ | $\beta=-2$ | $\beta=-1.5$ | $\delta=0.1$ | $\delta=0.25$ | $\delta=0.5$ |
| BC | 0.69 | 0.06 | 0.04 | 0.52 | 0.34 | 0.25 | 0.06 | 0.02 |
| CC | 0.47 | 0.07 | 0.05 | 0.44 | 0.30 | 0.23 | 0.07 | 0.01 |
| DC | 0.46 | 0.04 | 0.05 | 0.37 | 0.24 | 0.20 | 0.08 | 0.04 |
| EC | 0.51 | 0.08 | 0.06 | 0.42 | 0.31 | 0.21 | 0.08 | 0.03 |
| SC | 0.51 | 0.08 | 0.06 | 0.40 | 0.32 | 0.21 | 0.08 | 0.03 |

**Supplementary Table S30**

Normalized consensus times for the Prison network with divergence of 0.2.

| | | Random Emergence | | | | | | |
|---|---|---|---|---|---|---|---|---|
| | No PP | PL-decay | | Exp-decay | | Social | | |
| | | α=-2 | α=-1.5 | β=-2 | β=-1.5 | δ=0.1 | δ=0.25 | δ=0.5 |
| | 1.00 | 0.29 | 0.20 | 0.91 | 0.73 | 0.50 | 0.24 | 0.08 |
| | | Leader-Centrality Emergence | | | | | | |
| | No PP | PL-decay | | Exp-decay | | Social | | |
| Centrality | | α=-2 | α=-1.5 | β=-2 | β=-1.5 | δ=0.1 | δ=0.25 | δ=0.5 |
| BC | 0.81 | 0.15 | 0.10 | 0.74 | 0.56 | 0.38 | 0.15 | 0.03 |
| CC | 0.83 | 0.15 | 0.09 | 0.73 | 0.57 | 0.39 | 0.15 | 0.04 |
| DC | 0.82 | 0.15 | 0.09 | 0.83 | 0.55 | 0.39 | 0.14 | 0.05 |
| EC | 0.90 | 0.15 | 0.10 | 0.79 | 0.58 | 0.39 | 0.15 | 0.03 |
| SC | 0.85 | 0.15 | 0.09 | 0.74 | 0.56 | 0.39 | 0.14 | 0.05 |

**Supplementary Table S31**

Normalized consensus times for the Zachary network with divergence of 0.2.

| | | Random Emergence | | | | | | |
|---|---|---|---|---|---|---|---|---|
| | No PP | PL-decay | | Exp-decay | | Social | | |
| | | α=-2 | α=-1.5 | β=-2 | β=-1.5 | δ=0.1 | δ=0.25 | δ=0.5 |
| | 1.00 | 0.27 | 0.21 | 0.83 | 0.64 | 0.40 | 0.19 | 0.08 |
| | | Leader-Centrality Emergence | | | | | | |
| | No PP | PL-decay | | Exp-decay | | Social | | |
| Centrality | | α=-2 | α=-1.5 | β=-2 | β=-1.5 | δ=0.1 | δ=0.25 | δ=0.5 |
| BC | 0.72 | 0.23 | 0.18 | 0.62 | 0.49 | 0.30 | 0.14 | 0.07 |
| CC | 0.72 | 0.23 | 0.18 | 0.61 | 0.48 | 0.30 | 0.14 | 0.06 |
| DC | 0.72 | 0.22 | 0.17 | 0.60 | 0.47 | 0.29 | 0.16 | 0.06 |
| EC | 0.70 | 0.22 | 0.17 | 0.59 | 0.48 | 0.31 | 0.15 | 0.05 |
| SC | 0.73 | 0.22 | 0.18 | 0.59 | 0.48 | 0.29 | 0.15 | 0.06 |

**Supplementary Table S32**

Normalized consensus times for the Colorado Springs network with divergence of 0.2.

|  | Random Emergence | | | | | | | |
|---|---|---|---|---|---|---|---|---|
|  | **No PP** | **PL-decay** | | **Exp-decay** | | **Social** | | |
|  |  | $\alpha=-2$ | $\alpha=-1.5$ | $\beta=-2$ | $\beta=-1.5$ | $\delta=0.1$ | $\delta=0.25$ | $\delta=0.5$ |
|  | 0.70 | 0.27 | 0.16 | 0.63 | 0.57 | 0.64 | 0.40 | 0.11 |
|  | Leader-Centrality Emergence | | | | | | | |
|  | **No PP** | **PL-decay** | | **Exp-decay** | | **Social** | | |
| **Centrality** |  | $\alpha=-2$ | $\alpha=-1.5$ | $\beta=-2$ | $\beta=-1.5$ | $\delta=0.1$ | $\delta=0.25$ | $\delta=0.5$ |
| BC | 1.00 | 0.11 | 0.07 | 0.86 | 0.63 | 0.50 | 0.20 | 0.04 |
| CC | 0.84 | 0.07 | 0.04 | 0.74 | 0.58 | 0.49 | 0.21 | 0.03 |
| DC | 0.81 | 0.13 | 0.09 | 0.73 | 0.56 | 0.43 | 0.20 | 0.05 |
| EC | 0.71 | 0.17 | 0.09 | 0.62 | 0.46 | 0.34 | 0.16 | 0.03 |
| SC | 0.91 | 0.17 | 0.11 | 0.85 | 0.64 | 0.49 | 0.24 | 0.06 |

**Supplementary Table S33**

Normalized consensus times for the Dolphins network with divergence of 0.2.

|  | Random Emergence | | | | | | | |
|---|---|---|---|---|---|---|---|---|
|  | **No PP** | **PL-decay** | | **Exp-decay** | | **Social** | | |
|  |  | $\alpha=-2$ | $\alpha=-1.5$ | $\beta=-2$ | $\beta=-1.5$ | $\delta=0.1$ | $\delta=0.25$ | $\delta=0.5$ |
|  | 0.84 | 0.24 | 0.18 | 0.76 | 0.59 | 0.41 | 0.19 | 0.08 |
|  | Leader-Centrality Emergence | | | | | | | |
|  | **No PP** | **PL-decay** | | **Exp-decay** | | **Social** | | |
| **Centrality** |  | $\alpha=-2$ | $\alpha=-1.5$ | $\beta=-2$ | $\beta=-1.5$ | $\delta=0.1$ | $\delta=0.25$ | $\delta=0.5$ |
| BC | 0.72 | 0.16 | 0.11 | 0.65 | 0.52 | 0.32 | 0.14 | 0.06 |
| CC | 0.92 | 0.16 | 0.10 | 0.82 | 0.60 | 0.40 | 0.14 | 0.05 |
| DC | 0.67 | 0.15 | 0.10 | 0.63 | 0.50 | 0.34 | 0.14 | 0.06 |
| EC | 0.94 | 0.19 | 0.12 | 0.84 | 0.63 | 0.41 | 0.17 | 0.05 |
| SC | 1.00 | 0.18 | 0.12 | 0.84 | 0.65 | 0.45 | 0.17 | 0.06 |

**Supplementary Table S34**

Normalized consensus times for the Galesburg network with divergence of 0.2.

| | Random Emergence | | | | | | | |
|---|---|---|---|---|---|---|---|---|
| | No PP | PL-decay | | Exp-decay | | Social | | |
| | | α=-2 | α=-1.5 | β=-2 | β=-1.5 | δ=0.1 | δ=0.25 | δ=0.5 |
| | 0.81 | 0.24 | 0.18 | 0.71 | 0.56 | 0.38 | 0.18 | 0.07 |
| | Leader-Centrality Emergence | | | | | | | |
| | No PP | PL-decay | | Exp-decay | | Social | | |
| Centrality | | α=-2 | α=-1.5 | β=-2 | β=-1.5 | δ=0.1 | δ=0.25 | δ=0.5 |
| BC | 0.80 | 0.20 | 0.15 | 0.70 | 0.56 | 0.36 | 0.15 | 0.06 |
| CC | 0.93 | 0.22 | 0.16 | 0.83 | 0.62 | 0.39 | 0.16 | 0.05 |
| DC | 0.94 | 0.22 | 0.16 | 0.83 | 0.64 | 0.39 | 0.16 | 0.06 |
| EC | 0.97 | 0.23 | 0.18 | 0.85 | 0.65 | 0.41 | 0.16 | 0.06 |
| SC | 1.00 | 0.24 | 0.17 | 0.84 | 0.64 | 0.40 | 0.17 | 0.06 |

**Supplementary Table S35**

Normalized consensus times for the HS network with divergence of 0.2

| | Random Emergence | | | | | | | |
|---|---|---|---|---|---|---|---|---|
| | No PP | PL-decay | | Exp-decay | | Social | | |
| | | α=-2 | α=-1.5 | β=-2 | β=-1.5 | δ=0.1 | δ=0.25 | δ=0.5 |
| | 0.76 | 0.23 | 0.17 | 0.65 | 0.51 | 0.37 | 0.20 | 0.08 |
| | Leader-Centrality Emergence | | | | | | | |
| | No PP | PL-decay | | Exp-decay | | Social | | |
| Centrality | | α=-2 | α=-1.5 | β=-2 | β=-1.5 | δ=0.1 | δ=0.25 | δ=0.5 |
| BC | 0.60 | 0.20 | 0.16 | 0.51 | 0.39 | 0.27 | 0.16 | 0.07 |
| CC | 0.99 | 0.21 | 0.16 | 0.83 | 0.60 | 0.34 | 0.15 | 0.06 |
| DC | 0.66 | 0.14 | 0.11 | 0.53 | 0.43 | 0.25 | 0.13 | 0.07 |
| EC | 0.99 | 0.23 | 0.17 | 0.82 | 0.62 | 0.35 | 0.15 | 0.05 |
| SC | 1.00 | 0.24 | 0.17 | 0.81 | 0.60 | 0.34 | 0.15 | 0.05 |

**Supplementary Table S36**

Normalized consensus times for the High Tech network with divergence of 0.2.

|  | Random Emergence | | | | | | | |
|---|---|---|---|---|---|---|---|---|
|  | No PP | PL-decay | | Exp-decay | | Social | | |
|  |  | α=-2 | α=-1.5 | β=-2 | β=-1.5 | δ=0.1 | δ=0.25 | δ=0.5 |
|  | 0.93 | 0.28 | 0.22 | 0.81 | 0.65 | 0.42 | 0.20 | 0.08 |
|  | Leader-Centrality Emergence | | | | | | | |
|  | No PP | PL-decay | | Exp-decay | | Social | | |
| Centrality |  | α=-2 | α=-1.5 | β=-2 | β=-1.5 | δ=0.1 | δ=0.25 | δ=0.5 |
| BC | 0.76 | 0.22 | 0.17 | 0.69 | 0.56 | 0.35 | 0.16 | 0.06 |
| CC | 0.94 | 0.25 | 0.18 | 0.81 | 0.64 | 0.40 | 0.17 | 0.05 |
| DC | 0.95 | 0.24 | 0.19 | 0.82 | 0.66 | 0.40 | 0.17 | 0.07 |
| EC | 0.97 | 0.27 | 0.20 | 0.84 | 0.67 | 0.41 | 0.18 | 0.07 |
| SC | 1.00 | 0.27 | 0.20 | 0.86 | 0.69 | 0.42 | 0.18 | 0.06 |

**Supplementary Table S37**

Normalized consensus times for the Math Method network with divergence of 0.2.

|  | Random Emergence | | | | | | | |
|---|---|---|---|---|---|---|---|---|
|  | No PP | PL-decay | | Exp-decay | | Social | | |
|  |  | α=-2 | α=-1.5 | β=-2 | β=-1.5 | δ=0.1 | δ=0.25 | δ=0.5 |
|  | 0.79 | 0.25 | 0.19 | 0.76 | 0.62 | 0.43 | 0.20 | 0.08 |
|  | Leader-Centrality Emergence | | | | | | | |
|  | No PP | PL-decay | | Exp-decay | | Social | | |
| Centrality |  | α=-2 | α=-1.5 | β=-2 | β=-1.5 | δ=0.1 | δ=0.25 | δ=0.5 |
| BC | 0.60 | 0.23 | 0.18 | 0.56 | 0.49 | 0.32 | 0.16 | 0.06 |
| CC | 1.00 | 0.26 | 0.18 | 0.90 | 0.68 | 0.43 | 0.19 | 0.07 |
| DC | 0.98 | 0.25 | 0.18 | 0.87 | 0.69 | 0.42 | 0.19 | 0.07 |
| EC | 0.96 | 0.25 | 0.18 | 0.85 | 0.71 | 0.42 | 0.20 | 0.06 |
| SC | 0.93 | 0.25 | 0.18 | 0.85 | 0.70 | 0.43 | 0.19 | 0.06 |

**Supplementary Table S38**

Normalized consensus times for the Sawmill network with divergence of 0.2.

|  | | Random Emergence | | | | | | |
|---|---|---|---|---|---|---|---|---|
|  | No PP | PL-decay | | Exp-decay | | Social | | |
|  |  | α=-2 | α=-1.5 | β=-2 | β=-1.5 | δ=0.1 | δ=0.25 | δ=0.5 |
|  | 0.96 | 0.23 | 0.16 | 0.87 | 0.66 | 0.43 | 0.19 | 0.06 |
|  | | Leader-Centrality Emergence | | | | | | |
|  | No PP | PL-decay | | Exp-decay | | Social | | |
| Centrality |  | α=-2 | α=-1.5 | β=-2 | β=-1.5 | δ=0.1 | δ=0.25 | δ=0.5 |
| BC | 0.63 | 0.17 | 0.12 | 0.59 | 0.46 | 0.33 | 0.14 | 0.05 |
| CC | 0.66 | 0.16 | 0.12 | 0.58 | 0.49 | 0.33 | 0.14 | 0.04 |
| DC | 0.64 | 0.15 | 0.10 | 0.60 | 0.48 | 0.33 | 0.14 | 0.04 |
| EC | 1.00 | 0.19 | 0.13 | 0.85 | 0.63 | 0.37 | 0.15 | 0.05 |
| SC | 0.75 | 0.17 | 0.12 | 0.65 | 0.51 | 0.35 | 0.14 | 0.04 |

**Supplementary Table S39**

Normalized consensus times for the Social3 network with divergence of 0.2.

|  | | Random Emergence | | | | | | |
|---|---|---|---|---|---|---|---|---|
|  | No PP | PL-decay | | Exp-decay | | Social | | |
|  |  | α=-2 | α=-1.5 | β=-2 | β=-1.5 | δ=0.1 | δ=0.25 | δ=0.5 |
|  | 0.94 | 0.37 | 0.29 | 0.83 | 0.71 | 0.50 | 0.28 | 0.13 |
|  | | Leader-Centrality Emergence | | | | | | |
|  | No PP | PL-decay | | Exp-decay | | Social | | |
| Centrality |  | α=-2 | α=-1.5 | β=-2 | β=-1.5 | δ=0.1 | δ=0.25 | δ=0.5 |
| BC | 0.90 | 0.35 | 0.27 | 0.83 | 0.70 | 0.43 | 0.23 | 0.08 |
| CC | 0.97 | 0.34 | 0.27 | 0.87 | 0.73 | 0.45 | 0.22 | 0.09 |
| DC | 0.90 | 0.36 | 0.27 | 0.87 | 0.72 | 0.45 | 0.22 | 0.08 |
| EC | 1.00 | 0.31 | 0.25 | 0.87 | 0.73 | 0.44 | 0.23 | 0.09 |
| SC | 0.94 | 0.33 | 0.24 | 0.83 | 0.70 | 0.44 | 0.21 | 0.11 |